\documentclass[useAMS,usenatbib]{mn2e}

\usepackage{amsmath} 
\usepackage{subfigure} 
\usepackage{epsfig}
\usepackage{appendix}
\usepackage{xspace}
\usepackage{multirow}
\usepackage[T1]{fontenc}
\usepackage{url}

\def\reff@jnl#1{{\rm#1\/}}
\def\aj{\reff@jnl{AJ}}         
\def\araa{\reff@jnl{ARA\&A}}      
\def\apj{\reff@jnl{ApJ}}        
\def\apjl{\reff@jnl{ApJ}}        
\def\apjs{\reff@jnl{ApJS}}       
\def\aap{\reff@jnl{A\&A}}        
\def\aapr{\reff@jnl{A\&A~Rev.}}     
\def\aaps{\reff@jnl{A\&AS}}       
\def\mnras{\reff@jnl{MNRAS}}      
\def\physrep{\reff@jnl{Physics Reports}}
\def\prd{\reff@jnl{Phys.Rev.D}}     
\def\prl{\reff@jnl{Phys.Rev.Lett}}   
\def\pasp{\reff@jnl{PASP}}       
\def\pasj{\reff@jnl{PASJ}}       
\def\nat{\reff@jnl{Nature}}       

\def\Sref#1{Section~\ref{#1}\xspace}

\def\Fref#1{Figure~\ref{#1}\xspace}

\def\Tref#1{Table~\ref{#1}\xspace}
\def\Eref#1{Equation~\ref{#1}\xspace}
\def\erefa#1{Equations~\ref{#1}\xspace}
\def\erefb#1{\ref{#1}\xspace}
\def\Aref#1{Appendix~\ref{#1}\xspace}
\def\eg{{\it e.g.}}

\def\phosim{{\sc PhoSim}\xspace}
\def\imcat{{\sc imcat}\xspace}
\def\psfent{{\sc psfent}\xspace}
\def\memsys{{\sc MemSys4}\xspace}

\def\lensfit{{\sc LensFit}\xspace}

\def\pr{{\rm Pr}}
\def\sigmaePSF{\sigma[\varepsilon_{\rm PSF}]} 
\def\sigmasyssq{{\tilde{\sigma}_{\rm sys,PSF}}^{2}}

\usepackage[usenames]{color}
\usepackage{graphicx}

\newcommand{\chihway}[1]{{\textcolor{black}{#1}}}
\def\kipac{KIPAC, Stanford University, 452 Lomita Mall, 
Stanford, CA 94309, USA}

\def\oxford{Department of Physics, University of Oxford, 
 Keble Road, Oxford, OX1 3RH, UK}
\def\ssl{Space Sciences Laboratory, University of California, 
 Berkeley, CA 94720, USA}
\def\purdue{Department of Physics, Purdue University, 
 West Lafayette, IN 47907, USA}
\def\uw{Department of Astronomy, University of Washington, 
Seattle, WA 98195}
\def\cavendish{Cavendish Laboratory, Cambridge University, 
Madingley Road, Cambridge, CB2 3RF, UK}

\title[Atmospheric PSF Interpolation for Weak Lensing]
{Atmospheric PSF Interpolation for Weak Lensing \\
in Short Exposure Imaging Data}

\author[C.~Chang et al.]
{C.~Chang,$^{1}$\thanks{E-mail: chihway@slac.stanford.edu}
P.~J.~Marshall,$^{2}$
J.~G.~Jernigan,$^{3}$
J.~R.~Peterson,$^{4}$
S.~M.~Kahn,$^{1}$
\newauthor
S.~F.~Gull,$^{5}$
Y.~AlSayyad,$^{6}$ 
Z.~Ahmad,$^{4}$
J.~Bankert,$^{4}$ 
D.~Bard,$^{1}$
A.~Connolly,$^{6}$ 
\newauthor
R.~R.~Gibson,$^{6}$ 
K.~Gilmore,$^{1}$
E.~Grace,$^{4}$ 
M.~Hannel,$^{4}$
M.~A.~Hodge,$^{4}$ 
L.~Jones,$^{6}$ 
\newauthor
S.~Krughoff,$^{6}$ 
S.~Lorenz,$^{4}$
S.~Marshall,$^{1}$ 
A.~Meert,$^{4}$
S.~Nagarajan,$^{4}$ 
E.~Peng,$^{4}$ 
\newauthor
A.~P.~Rasmussen,$^{1}$
M.~Shmakova,$^{1}$ 
N.~Sylvestre,$^{4}$
N.~Todd,$^{4}$ 
M.~Young$^{4}$ \\ \\
$^{1}$\kipac\\
$^{2}$\oxford\\
$^{3}$\ssl\\
$^{4}$\purdue \\
$^{5}$\cavendish \\
$^{6}$\uw\\}

\begin{document}

\date{Accepted, Received; in original form }

\pagerange{\pageref{firstpage}--\pageref{lastpage}} \pubyear{2011}

\maketitle

\label{firstpage}

\begin{abstract}
\chihway{A main science goal for the Large Synoptic Survey Telescope (LSST) is to 
measure the cosmic shear signal from weak lensing to extreme accuracy. One 
difficulty, however, is that with the short exposure time ($\simeq$15 seconds) 
proposed, the spatial variation of the Point Spread Function 
(PSF) shapes may be dominated by the atmosphere, in addition to optics errors.
While optics errors mainly cause the PSF to vary on angular scales similar or 
larger than a single CCD sensor, the atmosphere generates stochastic structures 
on a wide range of angular scales. It thus becomes a challenge to infer the multi-scale, 
complex atmospheric PSF patterns by interpolating the sparsely sampled stars in the field.} 
In this paper we present a new method, \psfent, for interpolating the PSF shape parameters, 
based on reconstructing underlying shape parameter maps with a multi-scale maximum 
entropy algorithm. We demonstrate, using images from the LSST Photon Simulator, 
the performance of our approach relative to a 5th-order polynomial fit (representing the 
current standard) and a simple boxcar smoothing technique. Quantitatively, \psfent predicts 
more accurate PSF models in all scenarios and the residual PSF errors are spatially less 
correlated. This improvement in PSF interpolation leads to a factor of 3.5 lower 
systematic errors in the shear power spectrum on scales smaller than $\sim13'$, 
compared to polynomial fitting. We estimate that with \psfent and for stellar 
densities greater than $\simeq$1$/{\rm arcmin}^{2}$, the spurious shear correlation 
from PSF interpolation, after combining a complete 10-year dataset from LSST, is lower 
than the corresponding statistical uncertainties on the cosmic shear power spectrum, even 
under a conservative scenario.

\end{abstract}



\section{Introduction}
\label{sec:Introduction}

Gravitational lensing is the physical phenomenon where gravitational
fields perturb the trajectory of light rays and therefore distort observed
images \citep{1992LNP...406..196S}. In particular, the study of weak 
gravitational lensing involves measuring the statistical properties of an ensemble 
of distorted galaxy images \citep[see \eg][]{2001PhR...340..291B}. Weak 
gravitational lensing is, in principle, one of the most powerful probes of dark 
matter and dark energy. By measuring, at different redshifts, the statistical 
distortion of background galaxies due to large scale cosmic structures -- the 
``cosmic shear'' -- it is possible to place extremely tight constraints on the nature 
of dark energy \citep{1997ApJ...484..560J, 1999ApJ...514L..65H}. 

\chihway{
For observations to date, the accuracy of cosmic shear measurements has been
mostly limited by the statistical variation of random galaxy shapes in the relatively
small sky areas studied \citep{2006ApJ...647..116H, 2006A&A...452...51S, 
2006Msngr.126...19H, 2006ApJ...644...71J, 2007A&A...468..859H, 
2007MNRAS.381..702B,2010A&A...516A..63S}.} However, in future wide-field weak
lensing surveys such as those planned with
the Dark Energy Survey,\footnote{\url{http://www.darkenergysurvey.org/}} 
LSST\footnote{\url{http://www.lsst.org/}} \citep{2008arXiv0805.2366I}, 
and Euclid,\footnote{\url{http://sci.esa.int/euclid}}
extremely large datasets will greatly reduce the statistical errors, making these 
experiments systematics-limited.

A major source of systematic error in weak lensing comes from our incomplete
knowledge of the PSF \citep{2008A&A...484...67P}. To account for the effect of
the PSF on observed galaxy images, we need a model for it at
every galaxy position; these models can be constrained by images of stars,
which provide noisy estimates of the PSF shape that are more sparsely 
distributed than the galaxies. This is the ``PSF interpolation problem''.
In an earlier paper \citep[][hereafter C12]{C12}, we quantified 
the spatial variation in the PSF shapes for a typical LSST 15-second exposure
due to various physical effects, such as optics misalignments and atmospheric
turbulence. We found that, although most of the PSF anisotropy due to
instrumental effects varies smoothly over the field of view, the atmospheric
turbulence can generate PSF spatial variation on a wide range of scales in
these short exposures, with patterns that do not repeat over time. This
poses a new PSF interpolation challenge quite different from that faced by 
previous studies, which relied on images with longer exposure times and/or 
contained large instrumental effects that dominate the errors.

These atmospheric features have been observed in short exposure ($\sim10$ 
seconds) images by, for example, \citet{2005ApJ...632L...5W}
and \citet[][hereafter H12]{2012MNRAS.421..381H}; 
\chihway{in particular, H12 pointed out that such high frequency, turbulence-induced 
spatial PSF variations in single short exposures may lead to systematic errors 
in shear measurements at levels that cannot be ignored for future weak lensing surveys, 
were existing PSF interpolation techniques to be used.}

In weak lensing analyses to date, the scheme used most often to interpolate
the PSF between sparsely sampled stars has been to fit a low order
two-dimensional spatial polynomial function to the stars' shape parameters.
\chihway{ \citet{2002A&A...393..369V} claimed that 2nd-order polynomials are
sufficient to model the PSF anisotropy variation across a typical CCD sensor 
of size $\sim10'$, while other studies identified possible drawbacks of simple 
polynomial fitting and more sophisticated models have been suggested
\citep[\eg][]{2002ASPC..283..193M, 2004MNRAS.347.1337H, 2005A&A...429...75V, 
2012MNRAS.419.2356B}. In general, being dominated by instrumental effects, 
which primarily generate large-scale, smooth features \citep{2004astro.ph.12234J, 
2007ApJS..172..203R, 2010A&A...516A..63S}, the PSF anisotropy in long exposure 
data can be modelled reasonably well using low order interpolation methods, even 
though these patterns are only sparsely sampled by the relatively low density of stars 
in the field.} However, for future synoptic wide-field surveys, such as LSST, where 
high cadence imaging is required, short exposures are inevitable. It is 
therefore important to re-examine the traditional PSF interpolation techniques, 
and develop new interpolation algorithms that are better suited for these data.

A deliberate effort is being made to study in detail the PSF patterns for 
LSST by carrying out end-to-end, photon-by-photon image simulations using
the LSST Photon Simulator \citep[\phosim;][]{P12,P10,2010SPIE.7738E..53C}.
\footnote{\url{http://lsst.astro.washington.edu/}} 
The ray-tracing procedure includes models for the instrument response,
telescope optics, and, most importantly for our purposes, the atmosphere.
For a full description of the atmospheric model and quantitative comparison 
of the model against real data, we refer the reader to \citet[][hereafter P12]{P12}. In
this work, our focus is on studying the PSF interpolation problem, for which
we require simulated data containing {\it realistic atmospheric PSFs, with a range 
of strengths and spatial scales}. The \phosim images meet these criteria, as we 
demonstrated in P12 by comparing the relevant PSF characteristics in our simulated 
images with those seen by H12 in short exposure images taken by the MegaCam 
wide-field camera (1 degree$^{2}$) on the Canada-France-Hawaii Telescope 
(CFHT). In this work we therefore test our new algorithm on \phosim simulated 
images, for which we know the underlying PSF spatial variation.

Working with simulated images is vital for this particular task, since existing 
data that are suitable for our tests are very limited -- our tests require wide-field, 
short-exposure images with high stellar densities, and taken under a wide 
range of atmospheric conditions. Simulations, with sufficiently high fidelity, allow 
us to test the absolute, as well as relative, accuracy of our algorithm in a more 
controlled fashion and with higher statistics. Our \phosim approach can be seen 
as the next step beyond that taken by \citet{2012MNRAS.419.2356B}, who re-sampled 
PSF patterns observed in Subaru images. Here, we use \phosim to generate large 
numbers of predicted LSST images with the relevant exposure time, 15 seconds -- 
effectively amplifying the data taken with CFHT for a more rigorous testing.

The primary aim of this paper is to introduce a new PSF interpolation technique, 
and test it under controlled conditions. In both real and simulated PSF patterns we 
observe structure due to the atmosphere on many different angular scales: this 
motivates us to model the underlying anisotropy maps using a range of different-sized 
smoothing kernels. We infer the pixel values of these underlying maps from the noisy, 
sparsely sampled stellar shape data given a non-committal entropic prior. We 
quantitatively compare this new maximum entropy method with two other methods that 
represent the current standards, and investigate the performance of all three using 
high-fidelity simulations over a range of observing conditions. 

The structure of this paper is as follows. In \Sref{sec:Background}, we review briefly the 
relevant weak lensing theory and the PSF interpolation problem for weak lensing. We 
then describe, in \Sref{sec:PSFent}, our new PSF interpolation method and give 
arguments for why it is well-suited for the particular problem at hand. In 
\Sref{sec:Programme} we use simulated images to quantify the performance of our new 
method against two strawman PSF interpolation techniques. 
We also define the metrics that we use to quantify the performance of a given 
interpolation method. 
The results of this programme are presented in \Sref{sec:Results}. We then
discuss their implications in terms of the systematic errors in cosmic shear 
measurement for future weak lensing surveys and make suggestions for further 
improvements on \psfent in \Sref{sec:FurtherWork}. We conclude in 
\Sref{sec:Conclusions}.


\section{Weak lensing and PSF interpolation}
\label{sec:Background}

In the weak lensing regime, the effect of gravitational lensing is to add 
a small offset to the intrinsic ellipticity of each galaxy, where ellipticity is 
typically defined as a two-component complex ``spinor'', 
$\boldsymbol{\varepsilon} = \varepsilon_{1} + i\varepsilon_{2}$ 
\citep[see \eg][] {2002A&A...396....1S}. The resulting 
observed ellipticity is a noisy but unbiased estimator of the applied shear.
We then constructs certain statistics from these shear estimators to infer 
cosmology. One of the most popular statistics is the two-point ellipticity 
correlation function $\xi_{\pm}(\theta)$ for the ensemble of galaxies: 
\begin{equation}
 \xi_{\pm}(\theta)=\langle \varepsilon_{\rm t}(\theta_{0}) \varepsilon_{\rm t}
 (\theta_{0}+\theta) \rangle \pm \langle \varepsilon_{\times}(\theta_{0}) 
 \varepsilon_{\times}(\theta_{0}+\theta ) \rangle \;,
 \label{eq:cf}
\end{equation}
\noindent where the angle brackets indicate an average over 
galaxy pairs separated by $\theta$ (with one galaxy located at $\theta_{0}$) 
and the subscripts $t$ and $\times$ indicate an isotropised decomposition of 
$\boldsymbol{\varepsilon}$ along the line connecting a certain pair 
of galaxies. The shear correlation functions predicted from cosmology, 
compared with these observed galaxy ellipticity correlation functions, provide 
a route by which the cosmological parameters can be inferred.   

The major challenge in a ground-based weak lensing analysis is to account 
for the instrumental and atmospheric PSF contribution to the observed galaxy
shapes, such that these effects do not systematically contaminate the
shear signal one wishes to measure. This involves ``deconvolving''
(approximately) the PSF from the galaxy images, where the PSF at each 
galaxy's location is ``interpolated'' from the shapes of nearby stars. 
%

\chihway{A wide range of algorithms have been developed to model the shapes 
of the galaxies and stars, and to perform PSF deconvolution in the noisy data  
\citep[see \eg][for a summary of the various methods]{2006MNRAS.368.1323H, 
2007MNRAS.376...13M, 2010MNRAS.405.2044B, 2012MNRAS.423.3163K, 
2012arXiv1204.4096K}.} However, to date, the PSF interpolation problem has 
been taken to be of secondary importance, since the PSF ellipticity patterns in 
existing images appear to be largely instrumental in origin, somewhat repeatable, 
and well-modelled by smoothly-varying functions such as low-order polynomials. 
However, as we showed in P12 and C12, this may not be the case for future 
instruments such as LSST, which are specifically designed for weak lensing and 
have extremely tight requirements on the instrument-induced PSF anisotropy. In 
these circumstances, the atmospheric effects, which used to be subdominant to 
instrumental effects, now become one of the key components in determining the 
PSF shape and the PSF spatial variation. This implies that the PSF shapes may  
no longer be smoothly varying across the field and modelling the PSF variation 
with such assumptions may be problematic. In addition, since the 
atmospheric effects are more pronounced in short exposures, datasets such as 
LSST, which are composed of sets of multi-epoch short exposure images 
instead of one long exposure, may suffer more from the atmospheric effects 
in single exposures when constructing the PSF model from interpolation. 

\begin{figure*}
 \begin{center}
 \includegraphics[height=2.5in]{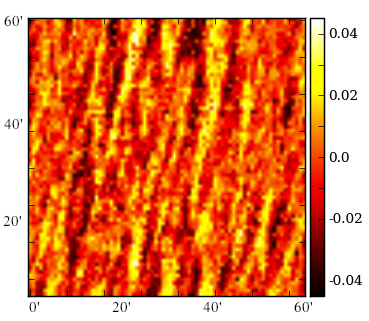} 
  \includegraphics[height=2.5in]{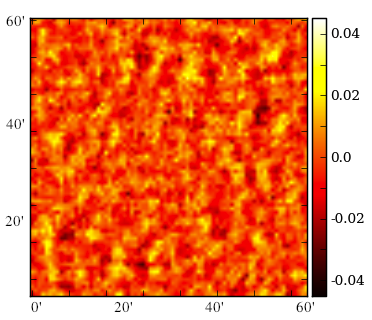} 
 \end{center}
 \caption{Examples of the residual PSF ellipticity ($\varepsilon_{1}$) patterns in $\sim$1 
 degree$^{2}$ sky regions, as observed in a dense stellar field with the CFHT MegaCam. 
 The exposure time for these images is 74 seconds. As described in H12, a 2nd-order 
 polynomial model has been fitted and subtracted from the raw stellar ellipticities to remove 
 the optics contribution.} 
 \label{fig:CFHT_maps}
\end{figure*}

To illustrate this PSF interpolation challenge, we show in \Fref{fig:CFHT_maps} 
two single-component ($\varepsilon_{1}$), model-subtracted stellar ellipticity maps 
from 74-second exposure images taken with the CFHT 
MegaCam\footnote{\url{http://www.cfht.hawaii.edu/Instruments/Imaging/Megacam/}} 
on a dense stellar field ($\sim7$ stars per arcmin$^{2}$). The two images used to construct 
\Fref{fig:CFHT_maps} were taken on the same patch of sky but in two different nights, which 
appear to have very different atmospheric conditions. The data in \Fref{fig:CFHT_maps} are 
taken from two of the catalogues used in H12, so we refer to their paper for further details of 
the dataset. Note that a 2nd-order polynomial was subtracted from the raw ellipticity 
measurements (as explained in H12) -- this accounts for most of the instrumental PSF 
contribution, but may also have removed some large scale 
atmospheric features. These images demonstrate that PSF spatial patters contain the 
characteristic high frequency structures from the atmosphere, which is the main motivation 
for our new PSF interpolation method. 

\section{\textbf{\textsc{PSFent}}: a multi-scale inferential interpolation method}
\label{sec:PSFent}

As seen in \Fref{fig:CFHT_maps}, the atmospheric PSF anisotropy patterns
can contain structure on a range of angular scales, with both patchy and
striped features. This motivates us to look for flexible functions with which
to model this spatial variability, which we can then fit to the sparsely sampled 
stellar PSF shape data in any given situation. 

\subsection{Interpolant model, and the likelihood function}
\label{sec:likelihood}

\chihway{Throughout this paper, we test the performance of our new interpolation 
method by interpolating two shape parameters: $\varepsilon_{1}$ and 
$\varepsilon_{2}$. In practice, a full weak lensing pipeline will require interpolation 
of several other shape parameters as well (\eg PSF size, matrix elements of the 
shear polarisability in KSB-type methods \citep{1995ApJ...449..460K} \textit{etc}). 
We do not do a full analysis on these other parameters, but demonstrate in 
\Aref{sec:R_test} that the our methodology is easily generalised to other parameters.}

\chihway{For the atmosphere-induced PSF, the two components of the ellipticity, 
$\varepsilon_{1}$ and $\varepsilon_{2}$, are expected to be independent from 
each other in each exposure, while both varying with similar amplitudes and spatial 
structures between different exposures. (We demonstrate this point with simulations 
in \Aref{sec:e1e2_corr}.) This is because the physical mechanism 
that generates these ellipticity values -- the refraction by turbulent cells -- does 
not have a preferred direction, but does have a certain magnitude and spatial power 
spectrum dictated by the local atmospheric parameters \citep{2007ApJ...662..744D}. 
As a result, we treat the 
two components of complex PSF ellipticity as independent fields with the same priors. 
These two fields, $\varepsilon_{1}(x,y)$ and $\varepsilon_{2}(x,y)$, are to be 
reconstructed from sparse, noisy, stellar shape data $\varepsilon^{\rm obs}_{1,k}$ and 
$\varepsilon^{\rm obs}_{2,k}$. Here, $k$ runs from 1 to the number of stars observed, 
$N_{star}$.} Casting the PSF interpolation problem as an image restoration
problem in this way allows us to properly take into account the observational
errors on the measured star shapes, and propagate those errors into
uncertainties on the interpolant. An iterative likelihood fit is performed: at each
step, the two ellipticity components of each stellar image are predicted from the 
model underlying ellipticity fields and compared to the measured stellar 
ellipticities. 

Both the predicted and observed data are inputs to the likelihood function.
Under the assumption of uncorrelated Gaussian stellar shape
uncertainties~$\sigma_{k}$, this can be written for \eg the first ellipticity
component as the following probability distribution (PDF) for the data:
\begin{align}
  \pr(\boldsymbol{\varepsilon}^{\rm obs}_{1}|&\boldsymbol{h}_1) 
  = \frac{1}{(2\pi)^{N_{star}/2} \prod_k \sigma_{1,k}} \notag \\
    &\times \exp{\left( -\frac{1}{2} 
            \sum_k \left[ \frac{\varepsilon^{\rm obs}_{1,k} -
                                \varepsilon_{1}(x_{k},y_{k};\boldsymbol{h}_1)}
                               {\sigma_{1,k}} \right]^2
      \right)},
  \label{eq:lhood}    
\end{align}
and likewise for $\varepsilon_2$. 

In \Eref{eq:lhood}, $\boldsymbol{h}_{1}$ is a parameter vector that represents 
the model. It is the components of this parameter vector (and its companion 
$\boldsymbol{h}_{2}$ for $\varepsilon_2$)
that we vary to fit the stellar shape data. We choose to parameterise the
flexible interpolation functions  $\varepsilon_{1}(x,y)$ and
$\varepsilon_{2}(x,y)$ with pixelated grids on the sky. We compute 
the predicted ellipticity at the $k$th star position,
$\varepsilon_{1}(x_{k},y_{k};\boldsymbol{h}_1)$, by linear interpolation between 
the neighbouring pixels: we choose the pixel scale of each grid on our maps such 
that each pixel contains approximately 1 target point, on average, such that the 
linear interpolation choice does not affect the final prediction. For our test data, we 
fix the model map sizes at $80\times80$ pixels for each ellipticity component.

Such free-form discretised functions like $\boldsymbol{h}_{1}$ and 
$\boldsymbol{h}_{2}$ have as many parameters to be inferred as
there are pixels in the grids. However, we would like to impose some
smoothness on these maps,  such that structure on a range of angular scales
can be predicted. We do this by constructing each map from a weighted sum of 
seven ``hidden'' maps, each convolved with a Gaussian ``Intrinsic Correlation
Function'' (ICF) of a different angular scale: the result is known as the
``visible'' map.  This procedure provides an efficient way of introducing
smooth, correlated structure on a variety of angular scales. In our notation, 
$\boldsymbol{h}$ stands for ``hidden.'' We therefore have $7\times80\times80 =
44,800$ hidden pixel values to vary during the fit, for each ellipticity
component. The convolutions with the
ICF kernels reduce the effective number of free parameters, but even so many
of these will still turn out not to be  constrained by the few hundred data
points in the field. The choice of prior PDF for the parameters in the
$\boldsymbol{h}$ is therefore important. 

\subsection{The entropic prior PDF}

\begin{figure}
  \begin{center}
  \begin{minipage}{0.9\linewidth}
    \begin{minipage}{\linewidth}
      \centering\includegraphics[width=0.3\linewidth]{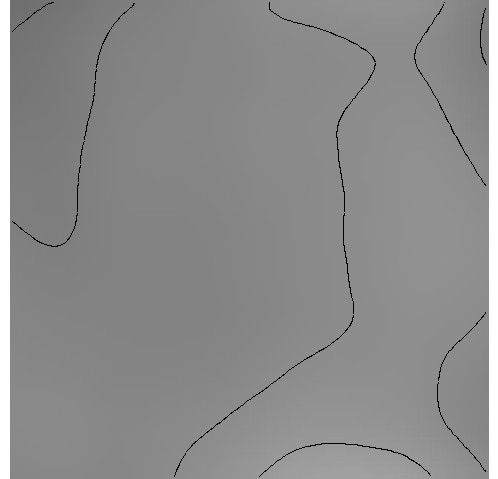}\hfill
      \raisebox{0.13\linewidth}{\large$\otimes$}
      \centering\includegraphics[width=0.3\linewidth]{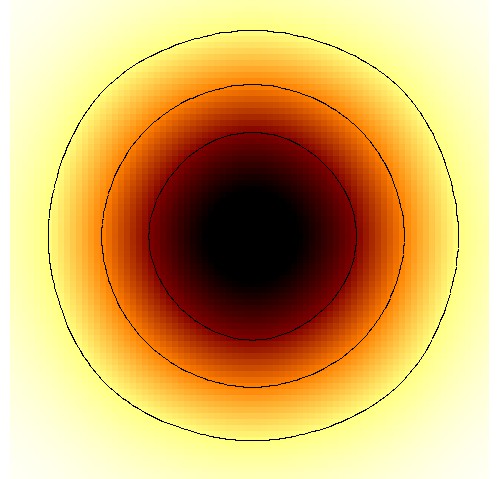}\hfill
      \raisebox{0.13\linewidth}{\large$=$}
      \centering\includegraphics[width=0.3\linewidth]{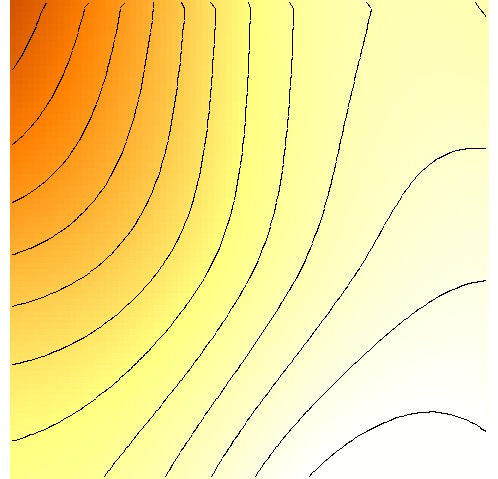}
    \end{minipage}
    \smallskip
    
    \begin{minipage}{\linewidth}
      \centering\includegraphics[width=0.3\linewidth]{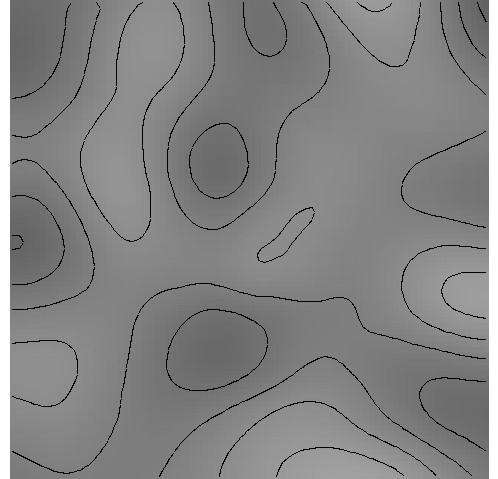}\hfill
      \raisebox{0.13\linewidth}{\large$\otimes$}
      \centering\includegraphics[width=0.3\linewidth]{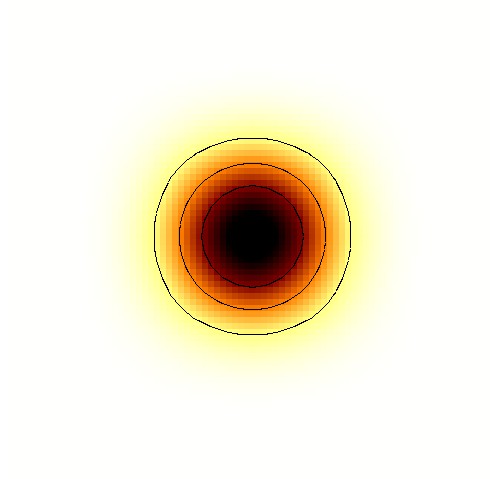}\hfill
      \raisebox{0.13\linewidth}{\large$=$}
      \centering\includegraphics[width=0.3\linewidth]{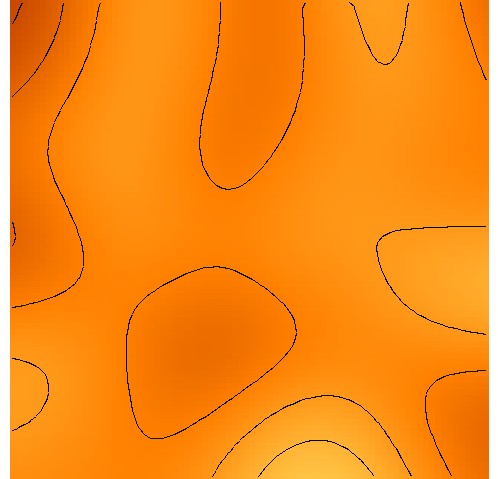}
    \end{minipage}
    \smallskip
    
    \begin{minipage}{\linewidth}
      \centering\includegraphics[width=0.3\linewidth]{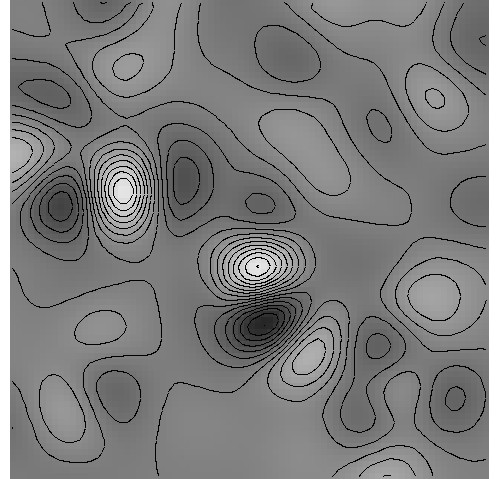}\hfill
      \raisebox{0.13\linewidth}{\large$\otimes$}
      \centering\includegraphics[width=0.3\linewidth]{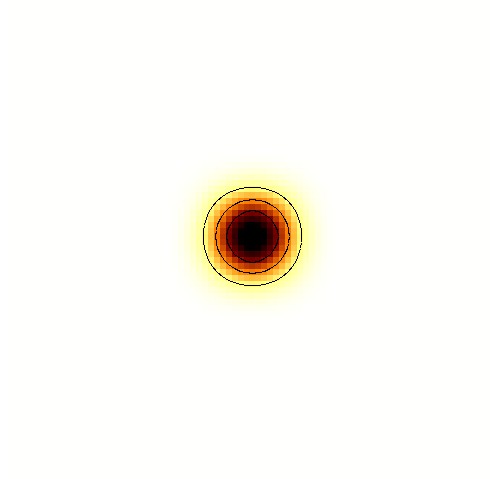}\hfill
      \raisebox{0.13\linewidth}{\large$=$}
      \centering\includegraphics[width=0.3\linewidth]{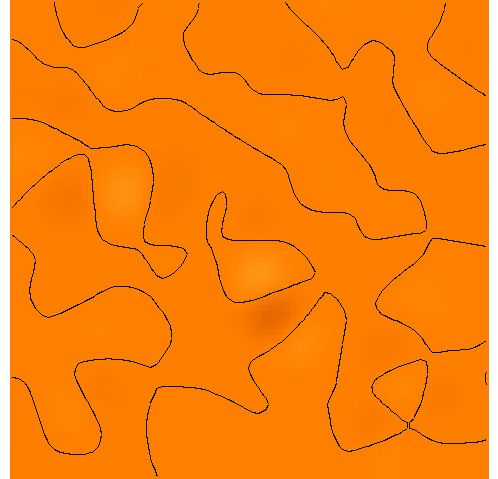}
    \end{minipage}
    \smallskip
    
    \begin{minipage}{\linewidth}
      \centering\includegraphics[width=0.3\linewidth]{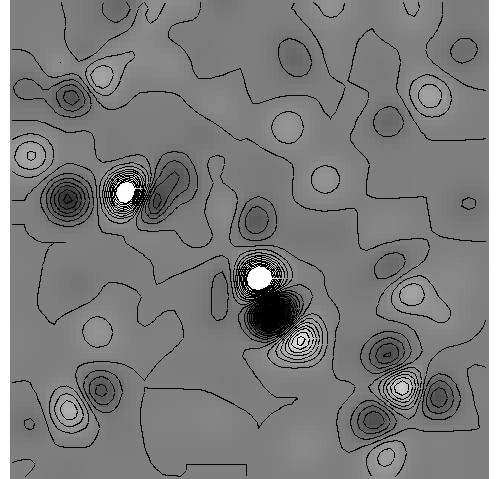}\hfill
      \raisebox{0.13\linewidth}{\large$\otimes$}
      \centering\includegraphics[width=0.3\linewidth]{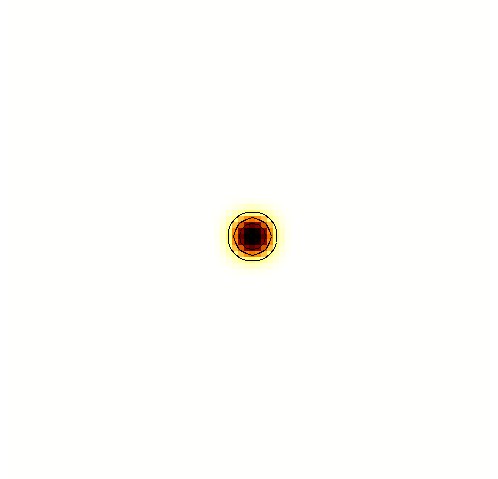}\hfill
      \raisebox{0.13\linewidth}{\large$=$}
      \centering\includegraphics[width=0.3\linewidth]{figs/hidden3.jpg}
    \end{minipage}
    \smallskip
    
    \begin{minipage}{\linewidth}
      \centering\includegraphics[width=0.3\linewidth]{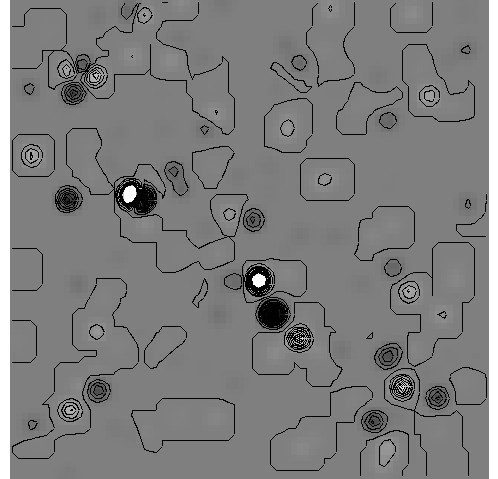}\hfill
      \raisebox{0.13\linewidth}{\large$\otimes$}
      \centering\includegraphics[width=0.3\linewidth]{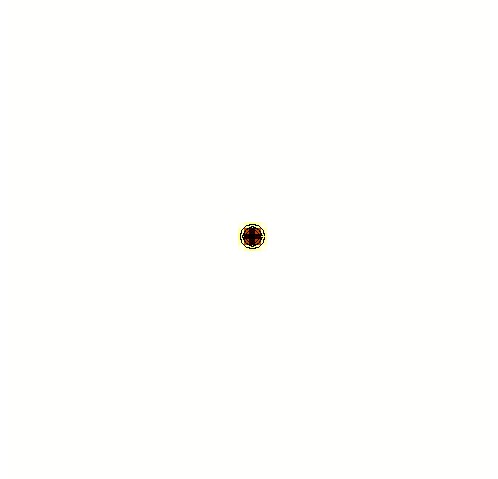}\hfill
      \raisebox{0.13\linewidth}{\large$=$}
      \centering\includegraphics[width=0.3\linewidth]{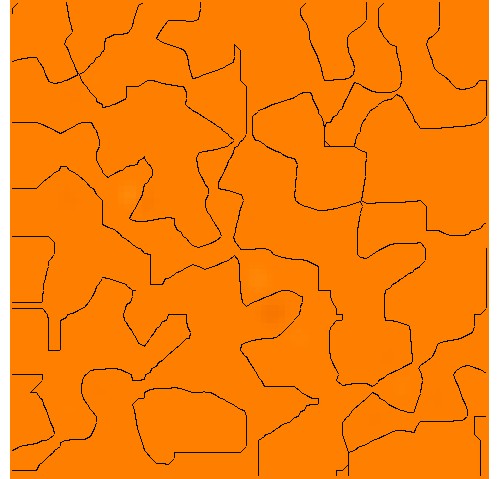}
    \end{minipage}
    \smallskip
    
    \begin{minipage}{\linewidth}
      \centering\includegraphics[width=0.3\linewidth]{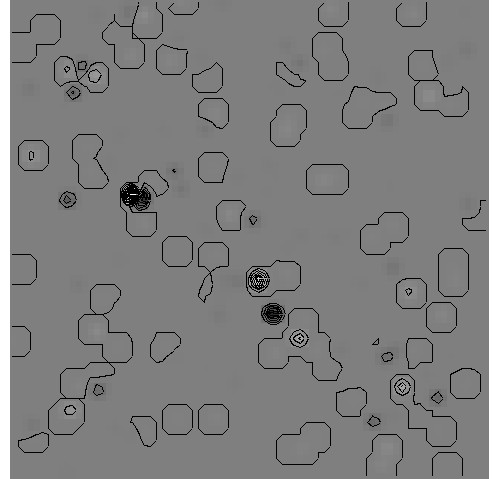}\hfill
      \raisebox{0.13\linewidth}{\large$\otimes$}
      \centering\includegraphics[width=0.3\linewidth]{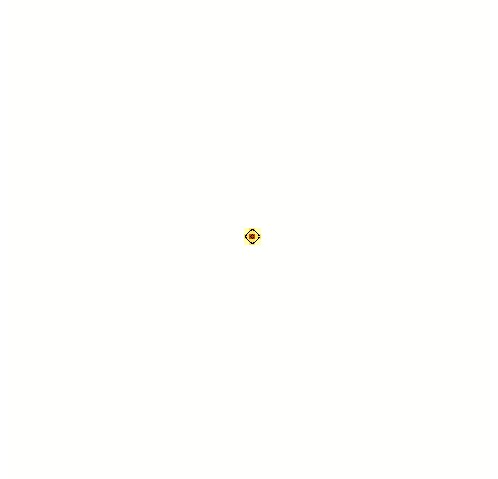}\hfill
      \raisebox{0.13\linewidth}{\large$=$}
      \centering\includegraphics[width=0.3\linewidth]{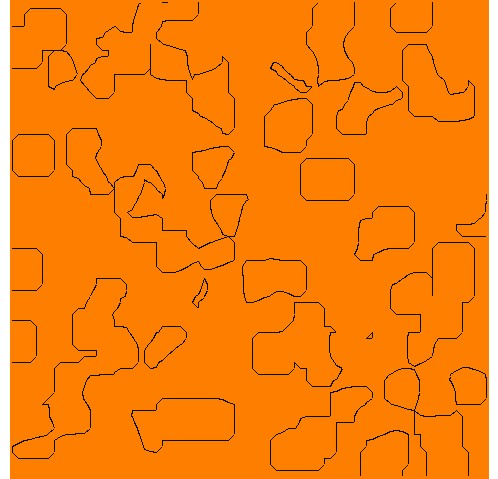}
    \end{minipage}
    \smallskip
    
    \begin{minipage}{\linewidth}
      \centering\includegraphics[width=0.3\linewidth]{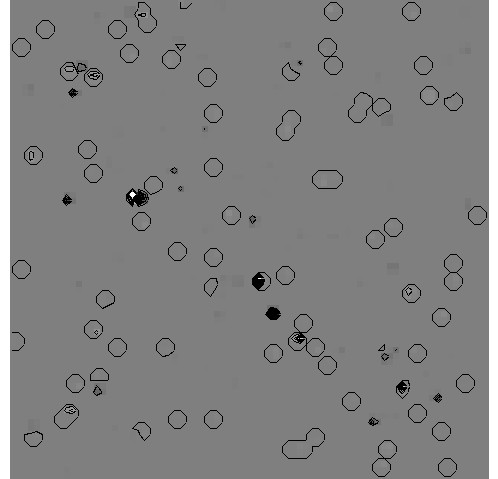}\hfill
      \raisebox{0.13\linewidth}{\large$\otimes$}
      \centering\includegraphics[width=0.3\linewidth]{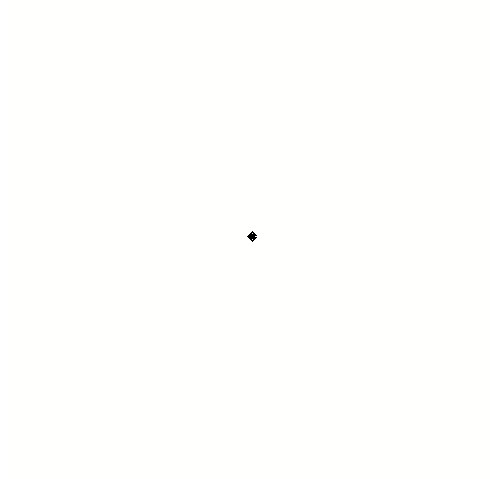}\hfill
      \raisebox{0.13\linewidth}{\large$=$}
      \centering\includegraphics[width=0.3\linewidth]{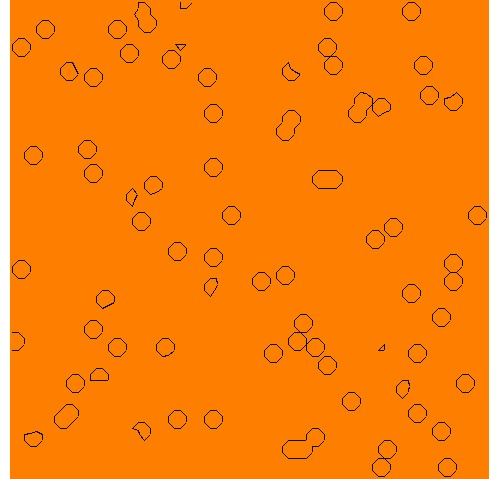}
    \end{minipage}
    \smallskip
    
    \begin{minipage}{\linewidth}
      \raggedleft \large{$+$} \rule{0.3\linewidth}{1pt}
    \end{minipage}
    \smallskip
    
    \begin{minipage}{\linewidth}
      \raggedleft\includegraphics[width=0.3\linewidth]{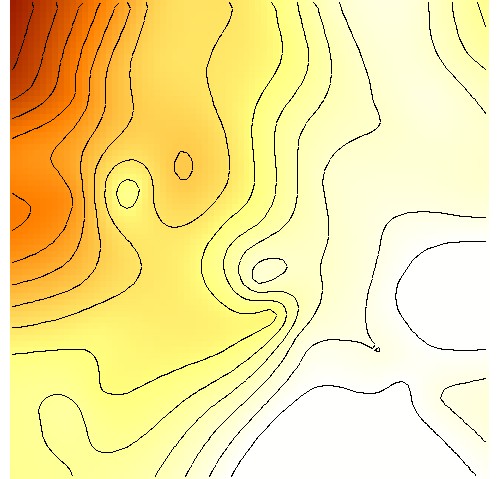}
    \end{minipage}
  \end{minipage}
  \end{center}
  \caption{Illustrating the \psfent multi-scale ellipticity map
  model. ``Hidden'' maps (left column, greyscale) are
  convolved with Gaussian kernels (ICFs) of exponentially-decreasing size 
  (middle column) to make seven component ``visible'' maps 
  (right, orange). These component maps are then weighted and summed to 
  make the final model ellipticity map, which can be interpolated linearly onto 
  any target position to predict the PSF shape there.}
  \label{fig:msmodel}
\end{figure}

We take the pixel values of the hidden images to be uncorrelated by
construction, and assign a {\it positive-negative entropic prior} for
them~\citep[][hereafter MHL04]{MHL04}. This has the effect of suppressing 
structure in the maps unless it is required by the data. In this way we give
the method plenty of flexibility to fit the data well, but regularise to
avoid  over-fitting.  Such a multi-scale maximum entropy method was first
used by \citet{1992ASPC...25..186W}, and is implemented in the publicaly
available \memsys code \citep{memsys}.  We illustrate the construction of a
multi-scale ellipticity component map in  \Fref{fig:msmodel}, showing  how a
range of different features on different angular scales can be modelled.

This model is similar in both essence and outcomes to one comprising a 
pixelated map and its ``\`a trous'' wavelet transform, as shown in some detail
by MHL04. This wavelet transform can also be written  as a set of convolutions; 
the implementation of MHL04 works well when  making maps of the
CMB temperature anisotropies, which also exhibit patchy  features on a range
of angular scales. Just like wavelet basis functions, the  Gaussian ICFs we
use have characteristic angular scales~$w_i$  that increase  approximately
exponentially: the smallest is a single map pixel ($w_0 = 1$),  while the
largest is approximately $w_6 = 2^{6} = 64$ pixels in size.  The ICFs are
normalised to unit volume.

The entropic prior PDF for a single hidden pixel value $h_i$ 
takes the following form
\begin{align}
\pr(h_i|m_i) &= \exp[{\alpha S(h_i)}]\; ,\\
      S(h_i) &= \psi_i - 2m_i - h_i \log{\left(\frac{\psi_i - h_i}{2 m_i}\right)} \;,
\label{eq:entropic_prior}
\end{align}
where $\psi_i = (h_i^2 + 4m_i^2)^{1/2}$. This distribution peaks at zero and
is symmetric. The
``regularisation constant'' $\alpha$ is a (nuisance) hyperparameter that 
parametrises the prior distribution. $\alpha$ is
inferred from the data via the Bayesian evidence internally by \memsys, and
controls the final importance of the prior  relative to the data. The
``model'' values $m_i$ (which we take to be constant over each hidden image) 
are also hyper-parameters, that determine the
ease  with which structure develops at each resolution scale: the smaller the
value of $m_i$, the stronger the suppression of features at that scale.

\subsection{Informing the prior}

At this point we might ask whether we can inject any more information into 
the problem by choosing values  of the $m_i$ to reflect the statistical
properties of the simulated atmospheric PSF  patterns. For a given pixel
value, the entropic prior has approximate width $m \approx \sigma_h^2/2$,
where $\sigma_{h}$ is the rms width of an approximating Gaussian
(MHL04). This suggests that a possible algorithm for assigning the
prior width at a particular resolution scale is to consider the variance of
pixel histograms of low noise ``true'' multi-scale ICF hidden ellipticity maps
at that scale. 

\begin{figure*}
  \begin{minipage}{0.48\linewidth}
    \centering\includegraphics[width=\linewidth]{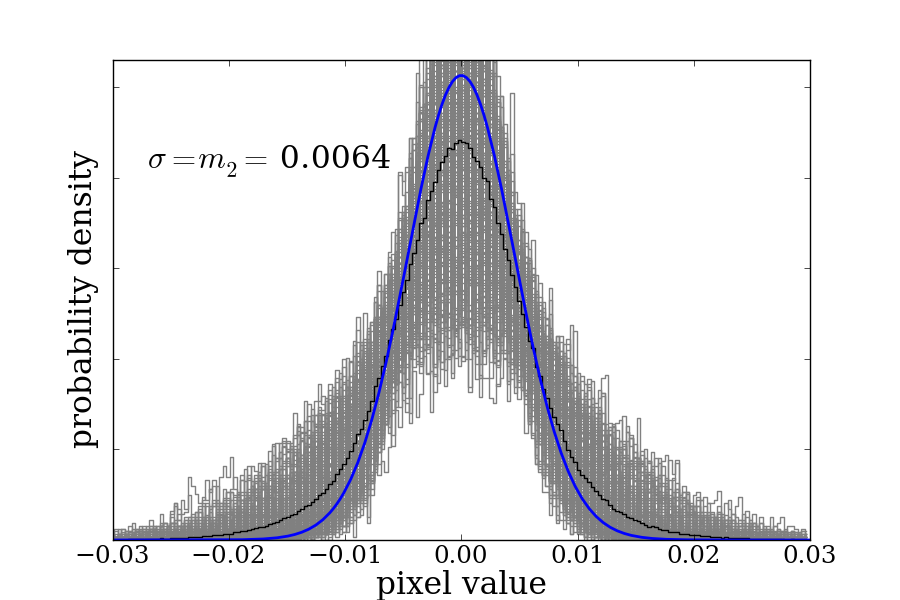}\hfill
  \end{minipage}
  \begin{minipage}{0.48\linewidth}
    \centering\includegraphics[width=\linewidth]{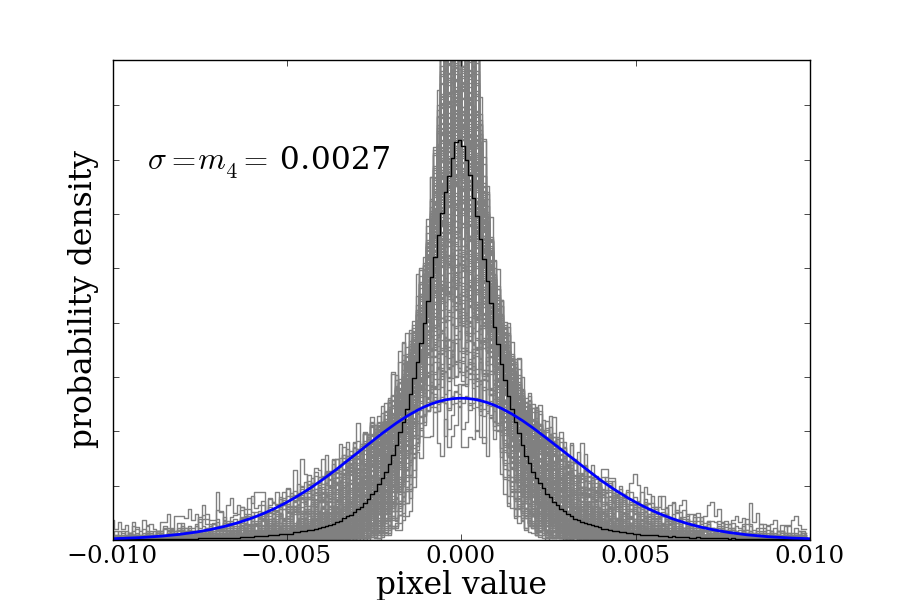}\hfill
  \end{minipage}
  \caption{Setting the entropic prior PDF on the map pixels. We show here in
  blue the entropic prior PDF (\Eref{eq:entropic_prior}) for a single pixel value $h$
  for two angular scales, 16 pixels (left) and 4 pixels (right). The grey curves are the 
  pixel histograms of simulated ellipticity component maps at that scale, and the
  black curve is the average of those histograms. The entropic prior peaks at
  zero as required, and has slightly longer tails than a Gaussian. While the
  atmospheric PSF ellipticity map histograms have even longer tails and sharper 
  peaks, especially on the smaller scales, by choosing
  the prior hyperparameter $m$ judiciously, the width of the histogram can be
  matched reasonably well. Note that the two plots have different scales on the 
  x-axis -- the histogram is much tighter for the smaller scales.}
  \label{fig:pixelhistograms}
\end{figure*}

To do this, we generate a special set of 100 simulated images 
with an ultra-high density of stars and low noise. \chihway{The range of PSF 
patterns in these simulations is consistent with the simulations used in the 
main analyses and is described in more detail in \Sref{sec:Programme_imsim}.} 
These unphysically dense 
star fields allow us to access the true PSF ellipticity pattern expected in the 
short exposures. The simulations are then run through \psfent to be effectively 
``decomposed'' into the 7 different scales, corresponding to the 7 ICFs. 
We generated one histogram for each of the 7 resolution scales that contain 
the pixel values ($\varepsilon_{1}$ and $\varepsilon_{2}$) in the hidden images 
for all 100 atmospheric PSF patterns on those scales. These histograms, as shown 
in \Fref{fig:pixelhistograms}, are more peaked, and have broader wings, than a 
Gaussian distribution. The entropic prior PDF (in blue), while still imperfect, is a slightly 
better approximation to these distributions. We found the rms widths of the average 
histograms at the different scales increase from the smallest scale to the largest scale 
as: $m_{i}$=[0.0003, 0.0002, 0.0027, 0.0050, 0.0064, 0.0089, 0.0188]. There is very 
little power on the smallest two scales and nonlinear increase from the remaining 
middle to large scales. We adopt these numbers as the width for the entropic priors 
throughout the rest of the paper. \chihway{We find that informing the priors in this way 
largely improves the accuracy of the PSF model constructed by \psfent, compared to 
using flat, non-informative priors. We also find that using the default prior set in \memsys 
(which was optimised for CMB temperature map reconstruction) results in well-behaved 
PSF models, although quantitatively worse than the priors constructed from realistic 
simulations described above\footnote{\chihway{We calculate the $\sigmaePSF$ and 
$\sigmasyssq$ values in the case of using \psfent with the \memsys priors to interpolate 
the PSF ellipticities for the same set of simulations in \Sref{sec:Results_patterns}. 
(See \Sref{sec:metrics} for the definition of these two statistics.) 
Quantitatively, using the \memsys priors increases $\sigmaePSF$ by $3\%$ and 
$\sigmasyssq$ by $29\%$ compared to using the priors derived from simulations.}}.}

\chihway{The standard deviation of the width of the grey histograms is approximately 
30 -- 40 $\%$ the mean width of the same histograms. This variation is due the 
variation in the atmospheric conditions in our simulations. For further optimisation of 
\psfent, we could also use a narrower range of weather conditions and carry out the same 
process multiple times to derive priors as a function of different atmospheric parameters 
such as seeing and wind speed. } 


As an aside, we note that the process of assigning prior widths for the different 
angular scales plays a very similar role to the assignment of the range 
parameter in the covariance function of the Gaussian process at the heart of a 
Kriging interpolation \citep{2012MNRAS.419.2356B}. The Kriging range parameter 
could also be derived by inspecting large numbers of high density simulated 
starfields, to capture the information present in the data-constrained atmospheric 
turbulence model in its most useful form.

\subsection{Estimating the posterior PDF}

The posterior PDF, \eg $\pr(\boldsymbol{h}_{j}|\{\varepsilon^{\rm obs}_{j}\})$, for
the hidden pixel values given the data can be approximated by a multivariate
Gaussian, centred at the maximum posterior point. The maximum posterior maps
provide our best estimates for the PSF shape parameters at any target point
in the field. Sample maps can be drawn from the posterior
PDF in order to provide approximate uncertainties on these estimated PSF
parameters. We find that 100 sample maps provide a sufficiently accurate
standard deviation map, which we use for the uncertainties on the predicted
PSF ellipticity estimates. This Gaussian approximation (including the maximum
of the posterior, the covariance matrix of the parameters, and the associated
evidence) is computed using the \memsys code, available on request from 
MaxEnt Data Consultants.\footnote{\url{http://www.maxent.co.uk}} Details of 
the implementation can be found in \citet{memsys}. 


\section{Simulation and analysis}
\label{sec:Programme}
  
\subsection{Simulations}
\label{sec:Programme_imsim}

As discussed in \Sref{sec:Introduction}, since the existing datasets are insufficient 
for us to perform a systematic test with \psfent, we depend on \phosim to generate 
a large number of simulated images. We refer to P12, \citet{P10} and 
\citet{2010SPIE.7738E..53C} for a complete description of \phosim and in specific the 
details of the atmospheric model. Here, to facilitate comparison with parallel PSF 
interpolation studies, we provide a very brief overview of the \phosim atmospheric 
model.

At the heart of the \phosim atmospheric model is a system of seven-layer frozen 
Kolmogorov screens \citep{Kolmogorov1941,Lane1992}. These screens are
constructed with density fluctuations obeying a full three-dimensional Kolmogorov 
energy spectrum of $E(k) \propto k^{-5/3}$, with values for the mean seeing, inner 
and outer turbulence scale assigned to each individual screen. All screens contain 
a wide range of turbulent structures, and these screens are carried by wind in 
different directions over the course of the exposure time. The distributions of wind 
speeds and atmosphere structure function parameters that \phosim uses are based 
on observed data taken near the LSST site at Cerro Pachon, Chile. In P12, we 
demonstrated that with this atmospheric model, the simulations from \phosim show 
sufficiently realistic atmosphere-induced PSF spatial variation for the purpose of 
weak lensing studies. Although \phosim was designed to simulate images from LSST 
in particular, the atmospheric model in \phosim generates PSF patterns qualitatively 
generic to most large aperture telescopes. The result of this study can thus be easily 
extended to estimate the performance of our PSF interpolation method on other 
instruments.    

\chihway{
Finally, in this paper, we follow the convention in C12 and simulate only the best 
50\% of the images in terms of image quality, where the median PSF size is 
$\sim0.7$" full-width-half-maximum (FWHM). This is based on the fact that in previous 
cosmic shear measurements, most of the cosmic shear information appear to come 
from these ``good'' images \citep{2006ApJ...647..116H}. That is, although one may be 
using all of the images, the bad images will be weighted low and thus the errors made 
in the PSF interpolation will also not be contributing as much to the final results.
The range of atmospheric parameters used in the this work is consistent with C12, 
which also allows us to make statements of the spurious shear correlation function in 
\Sref{sec:Result_ShearSys}.}

\subsection{Testing programme}

We use a mock stellar catalogue to generate realistic images of star fields. 
The catalogue is based on the model of \citet{2008ApJ...673..864J}, and 
contains a realistic population of stars in a typical LSST field with 
corresponding characteristics for each star. The average observed density 
of the population is $\simeq 1/{\rm arcmin}^{2}$, which corresponds to 
that expected at a galactic latitude of $|b|\sim60$; the equatorial coordinates 
of the portion of the star catalogue used for this baseline simulated field 
was $(1.5,+0.2)$ degrees. 

Two competing factors come into play in the PSF interpolation problem: (1)
the complexity of the PSF patterns, and (2) the number of stars available for
constructing a PSF model (or effectively, the galactic latitude). The more
complex the PSF pattern, or the fewer stars available to interpolate, the
more challenging it is to infer the PSF model from stars. The two effects are 
tested separately with the simulation and analysis pipeline described below.

We address (1) by generating 100 realisations of the atmospheric PSF 
patterns, and ``observing'' the mock star fields at these 100 different
``epochs.'' We generate these PSF realisations using the realistic 
distribution of atmospheric parameters described in P12. The median 
PSF size in our simulations is $\sim0.66''$ -- for LSST, this 
corresponds to approximately half of the exposures with the best image 
quality. Each atmosphere realisation creates a unique PSF pattern; 
simulating 100 independently corresponds to a low-cadence survey 
campaign where observations are well-separated in time. 

To investigate (2), we do not actually simulate star fields at different
galactic latitudes. Instead, we create an over-sampled stellar catalogue
based on the same stellar population used for (1), so that our \phosim input
catalogue contains higher stellar density than an ``average'' field, while
retaining the same signal to noise threshold in all ``detected'' star
catalogs. We then down-sample at the detection catalogue level to achieve
any desired stellar density used for interpolation -- which can then be
associated with a given galactic latitude. In this analysis, we consider 
stellar densities between the range 4 and 0.25 ${\rm arcmin}^{-2}$, which 
approximately covers the range of galactic latitudes $|b| >25$. As will be 
explained later, the stellar density quoted here is after a signal-to-noise ratio 
(SNR) cut, which eliminates very dim stars and noise peaks ($r>$23.5). In 
reality, harder cuts may be used to guarantee purity of the star sample. 

For each of the above scenarios we generate one image containing only 
the expected stars at their given positions, and a second image containing 
an ultra-dense grid of bright stars ($\sim50/{\rm arcmin}^{2}$) to sample the 
``true'' PSF pattern. Noise corresponding to a sky background level of 
22~mag/arcsec$^{2}$ was added to the star field images. All images were 
generated in the $r$ band for a single CCD sensor near the center of the LSST 
focal plane, which corresponds to a 13.6$' \times$13.6$'$ field on the sky.

The suite of simulated images were analysed using the same pipeline, 
in which stars were first detected using the Source Extractor 
\citep{1996A&AS..117..393B} package, and then catalogued using the \imcat
software developed by Nick
Kaiser.\footnote{\url{http://www.ifa.hawaii.edu/~kaiser/imcat/}} Shape
estimation was performed using the \imcat routine ``{\sc getshapes}''. We use
the output ``e[0]'' and ``e[1]'' as our representative measures of ellipticity 
($\varepsilon^{\rm obs}_{1,k}$ and $\varepsilon^{\rm obs}_{2,k}$ in 
\Sref{sec:PSFent}) and retained stars measured with \imcat parameter 
``$\nu$'' larger than 25 (equivalent to a signal-to-noise cut $\sim$13). 
In the absence of uncertainty estimates on the ellipticity components, we 
propagate $\nu$ as the stars' statistical weight.

\psfent, together with two other PSF modelling methods, polynomial fitting
and boxcar smoothing (details of our implementation of which are given in
\Aref{sec:methods}), were then applied to the detected stars. From each
method, the output was a list of predicted values of the PSF ellipticity at
the ultra-dense grid positions of the bright stars in the ``true PSF'' image.
These grid positions stand for background galaxy positions in a real
lensing analysis. In this way we were able to compare the output catalogues 
directly with the true underlying ellipticity maps. 


\subsection{Performance metrics}
\label{sec:metrics}

To quantitatively evaluate the performance of the different PSF interpolation 
techniques, we employ two performance metrics in this paper. The first metric, 
$\sigmaePSF$, is defined to be the root-mean-square of the PSF ellipticity 
model error: 
\begin{equation}
  \sigmaePSF=\sqrt{\langle \delta\varepsilon^{2}_{1}\rangle + 
  \langle \delta\varepsilon^{2}_{2}\rangle} \;.
\label{eq:sigma_e}
\end{equation}
\noindent where 
\begin{equation}   
   \delta\varepsilon_{i}=\varepsilon_{i,\rm model}-\varepsilon_{i,\rm true} \;.
\end{equation}  

The second metric, $\sigmasyssq$, is defined as the average amplitude of the 
two-point correlation function of the PSF ellipticity model errors in the scales of 
interest:
\begin{equation}
  \sigmasyssq=\frac{1}{\theta_{\rm max}-\theta_{\rm min}}
  \int_{\theta_{\rm min}}^{\theta_{\rm max}} |\xi^{s,\rm PSF}_{+}(\theta)|d\theta\; ,
 \label{eq:sigma_xi}
\end{equation}
\noindent where $\xi^{s,\rm PSF}_{+}(\theta)$ is the correlation function of 
$\delta\boldsymbol{\varepsilon}=\delta\varepsilon_{1}+i\delta\varepsilon_{2}$. 
The absolute value in the integrand prevents the anti-correlation regime canceling 
out some of the correlation signal. We use $\theta_{\rm min}=0.5'$ and 
$\theta_{\rm max}=13'$ in our single LSST CCD sensor ($13.6' \times13.6'$) simulations. 
This range is chosen to sample the correlated PSF model errors on the full sensor  
while eliminating small scales far below the average stellar separation and large 
scales that come close to the boundaries.
  
Note that $\sigmaePSF$ measures the level of the absolute errors in a certain PSF 
model, while $\sigmasyssq$ is a measure of the \textit{spatial correlation} of the PSF 
model errors, in addition to their absolute levels. This motivates us to take the 
ratio of these two contributions and define an auxiliary metric $F_{\rm sys}$:
\begin{equation}
F_{\rm sys}=\frac{\sigmasyssq}{\sigmaePSF^{2}}
\end{equation}
$F_{\rm sys}$ is a {\it relative} measure of the how spatially-correlated the residual ellipticities 
are, independent of the absolute magnitude of the residual ellipticities. As seen in later 
sections, this figure facilitates our comparison of different PSF interpolation methods. 

\subsection{Scaling with number of exposures}

\begin{figure*}
\begin{center}
  \subfigure[]{\includegraphics[scale=0.35]{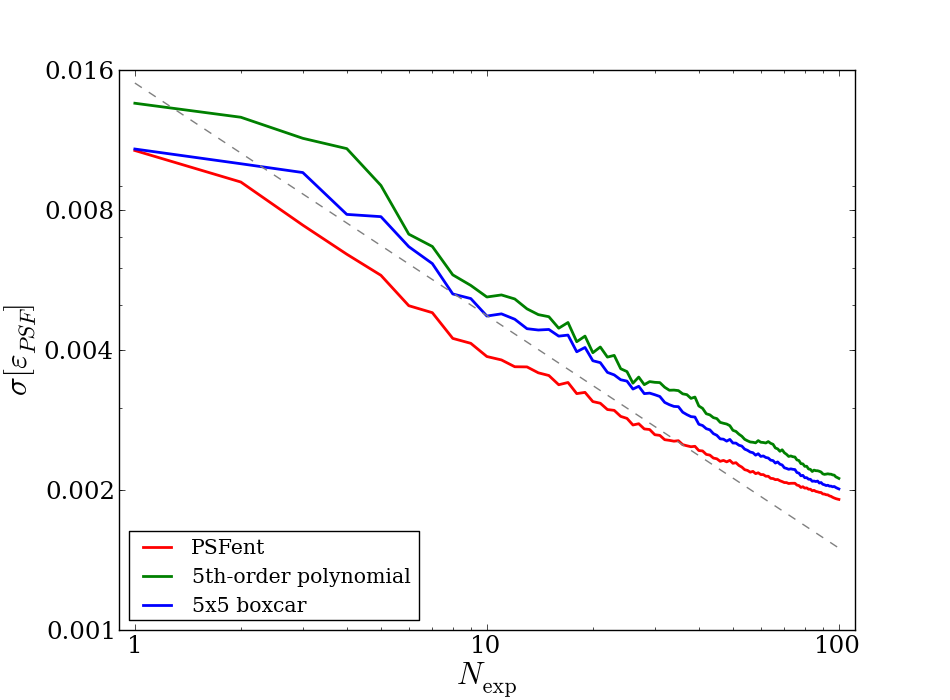} }
  \subfigure[]{\includegraphics[scale=0.35]{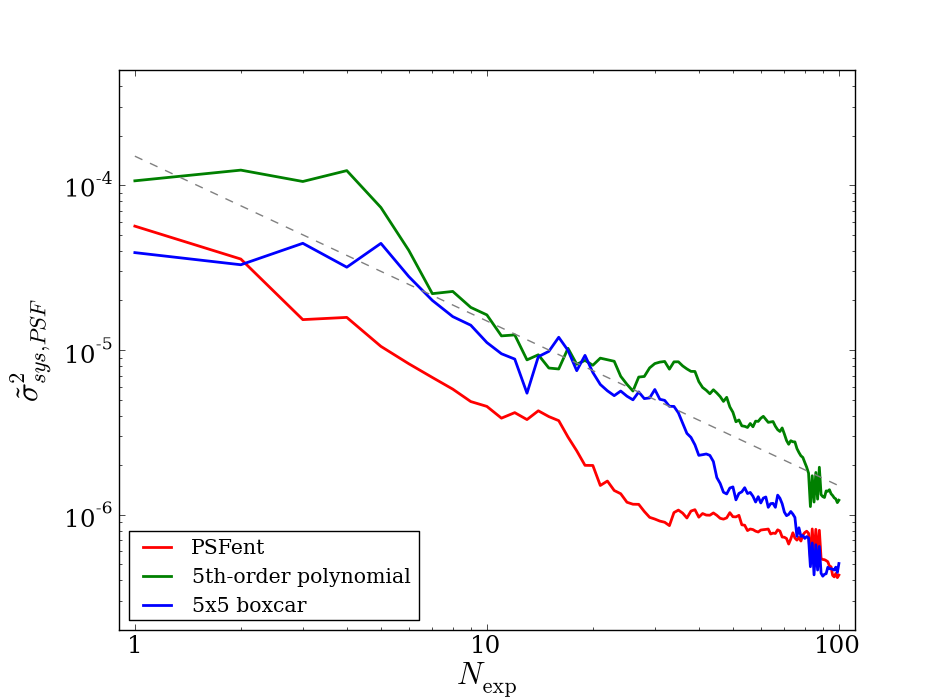} }
 \caption{(a) $\sigmaePSF$ and (b) $\sigmasyssq$ calculated as a function of the 
 number of exposures combined (averaged), $N_{\rm exp}$. We show in both panels 
 results for \psfent (red), 5th-order polynomial fitting (green) and 5$\times$5 boxcar 
 smoothing (blue). The grey dash line indicates the $1/\sqrt{N_{\rm exp}}$ and 
 $1/N_{\rm exp}$ slope with arbitrary normalisation for (a) and (b) respectively. All 
 three statistics in both plots roughly follow the $N_{\rm exp}$ scaling suggested by 
 the grey lines.}
 \label{fig:running_stats}
\end{center}
\end{figure*}

In the main analysis in this paper (\Sref{sec:Results}), we quantify the errors on 
the PSF ellipticity model for different interpolation techniques for a \textit{single 
LSST exposure}. In reality, one can suppress the systematic errors that are 
independent between frames when properly combining multiple exposures. It is 
the final \textit{combined} systematic error of all the data that impacts the 
cosmological constraints from weak lensing. 

When the multiple exposures are far separated in time, the atmospheric 
conditions are different and one expects the PSF patterns to be independent. 
Similarly, the errors on the PSF model are also expected to be independent. This 
suggests that when averaging the ellipticity measurement of the same galaxy over 
$N_{\rm exp}$ exposures, the ellipticity errors from the atmosphere are expected to 
reduced as $1/\sqrt{N_{\rm exp}}$ (corresponding to $\sigmaePSF$), and the 
correlation of these errors should drop by $1/N_{\rm exp}$ (corresponding to 
$\sigmasyssq$), or:

\begin{equation}
\sigmaePSF \propto \frac{1}{\sqrt{N_{\rm exp}}}\; ;
\label{eq:scale1}
\end{equation}
\begin{equation}
\sigmaePSF\propto \frac{1}{N_{\rm exp}} \; .
\label{eq:scale2}
\end{equation}

This can be nicely demonstrated by performing the following test: we take 
the 100 realisations of simulated atmospheric PSF and average the PSF model 
prediction at each position over the first $N_{\rm exp}$ frames.  $\sigmaePSF$ 
and $\sigmasyssq$ is then calculated for the ``average frame'' and plotted 
against $N_{\rm exp}$ in \Fref{fig:running_stats}. The two statistics scale with 
$N_{\rm exp}$ as expected in \erefa{eq:scale1} and \erefb{eq:scale2}, which 
confirms that the errors produced by all three interpolation methods used in this 
paper are stochastic over different exposures.  

We use the results in \Fref{fig:running_stats} to support arguments later in 
\Sref{sec:Result_ShearSys}, where we estimate the number of exposures needed 
to reach a certain level of PSF model accuracy.
 


\section{Results}
\label{sec:Results}


\subsection{Variation with PSF pattern}
\label{sec:Results_patterns}

\begin{figure*}
  \begin{center}
   \includegraphics[height=1.6in]{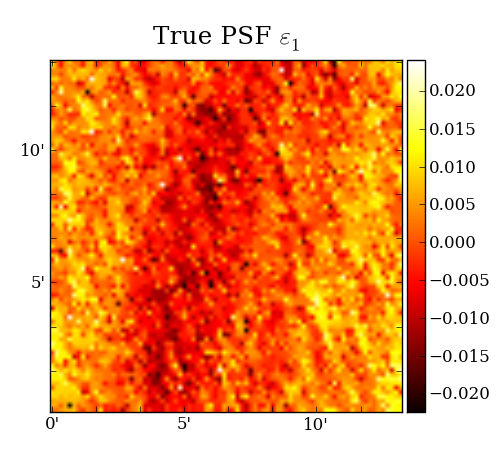}     
   \includegraphics[height=1.6in]{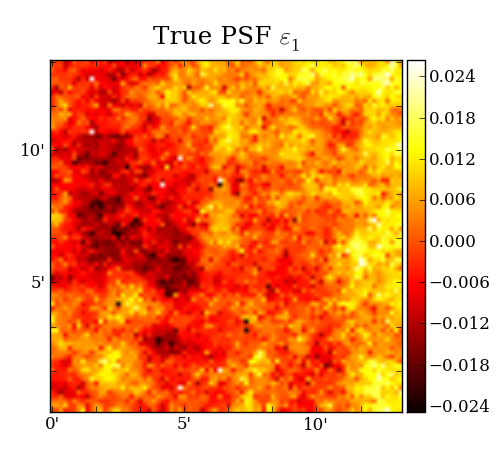}     
   \includegraphics[height=1.6in]{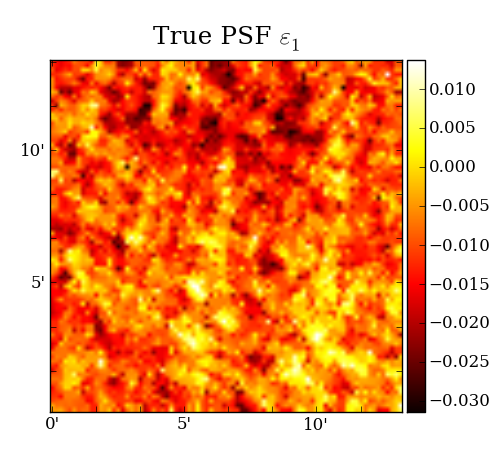}     \\ 
   \includegraphics[height=1.6in]{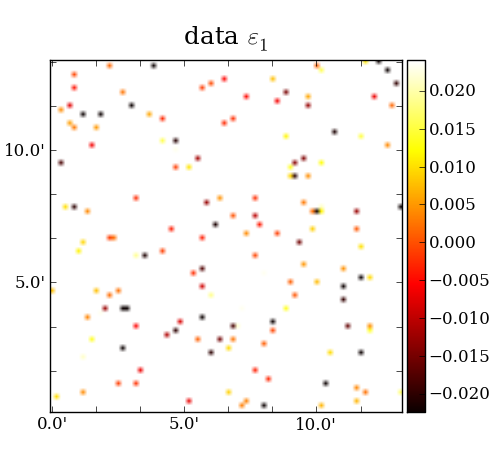}  
   \includegraphics[height=1.6in]{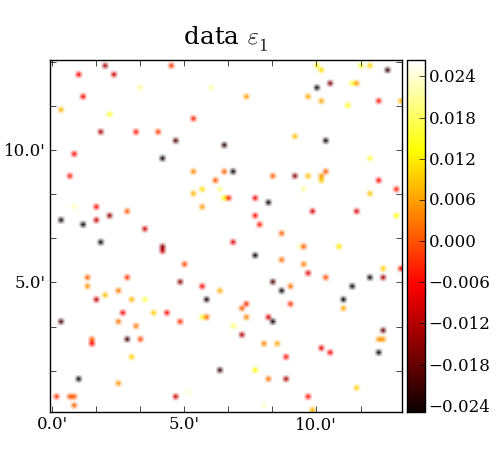}  
   \includegraphics[height=1.6in]{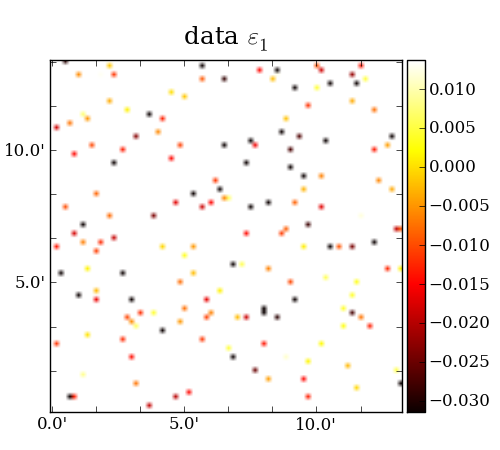}    \\ 
   \includegraphics[height=1.6in]{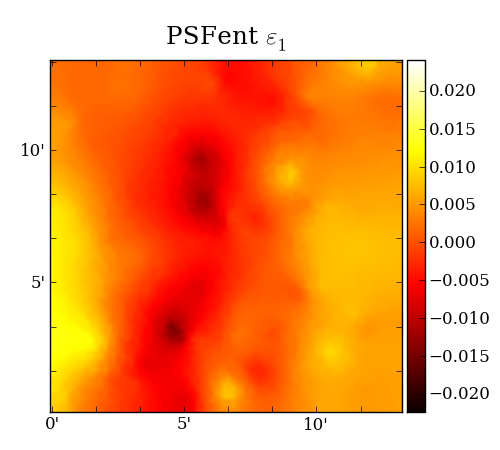}   
   \includegraphics[height=1.6in]{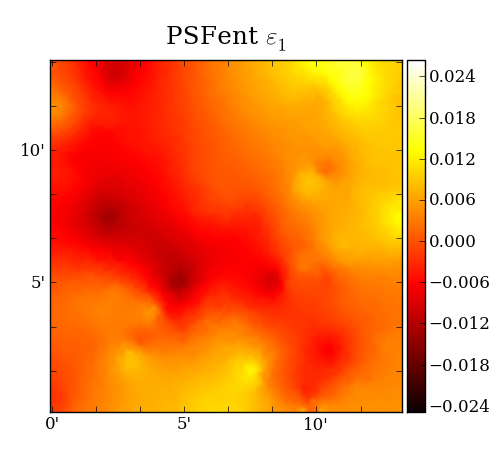}   
   \includegraphics[height=1.6in]{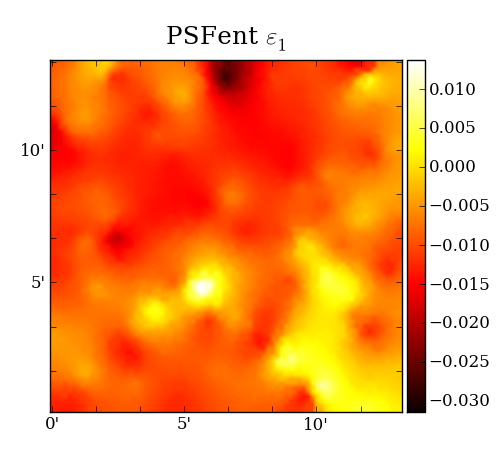}   \\ 
   \includegraphics[height=1.6in]{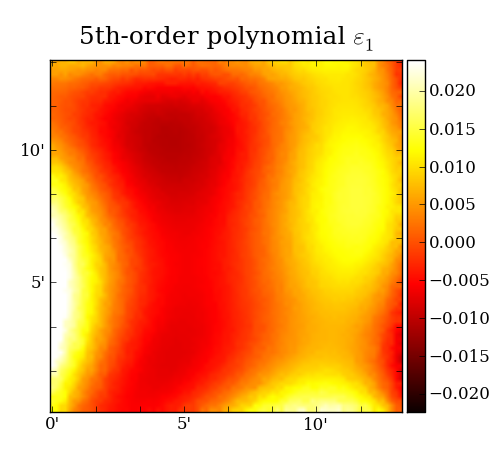}   
   \includegraphics[height=1.6in]{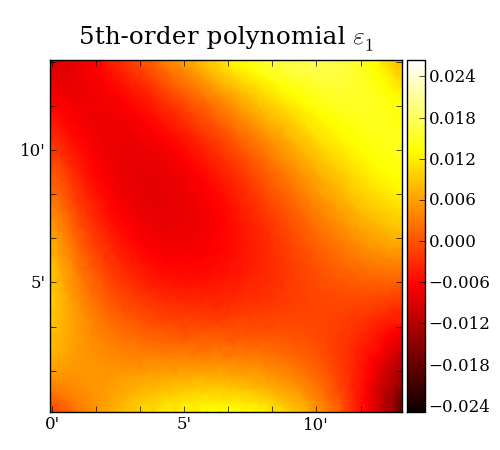}   
   \includegraphics[height=1.6in]{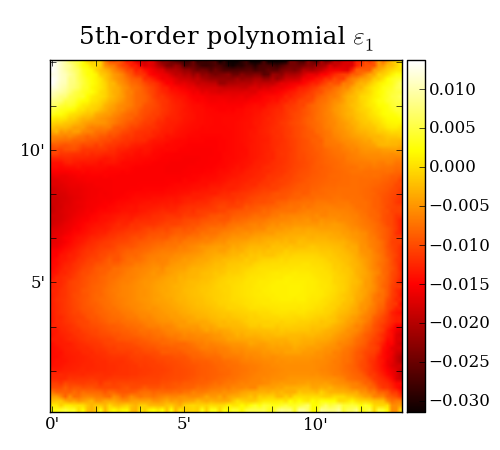}   \\ 
   \subfigure[]{\includegraphics[height=1.6in]{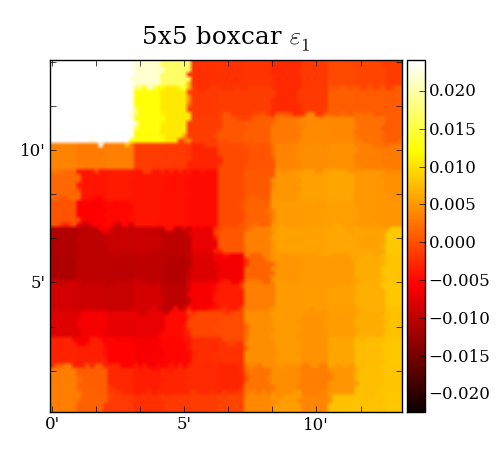} }  
   \subfigure[]{\includegraphics[height=1.6in]{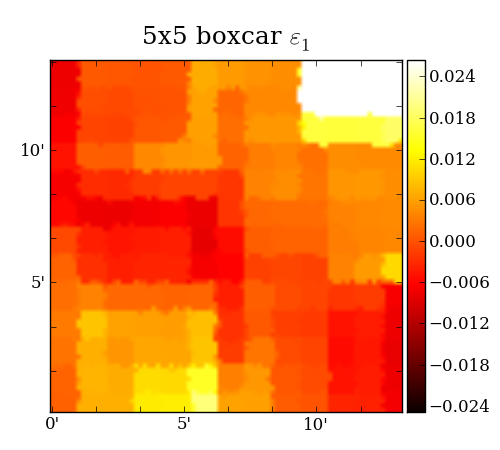} }  
   \subfigure[]{\includegraphics[height=1.6in]{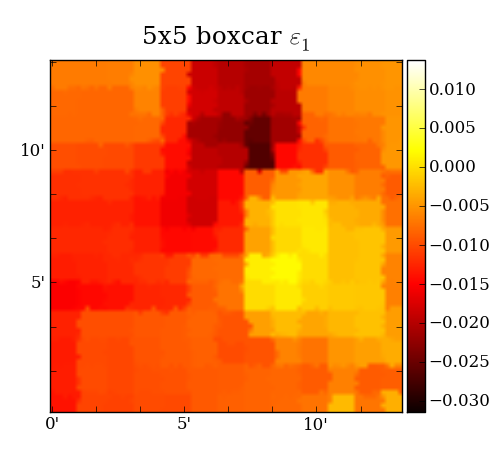} }  \\
 \end{center}
 \caption{Illustration of the short exposure PSF interpolation problem,
 and the  performance of different interpolation methods when we have very 
 different PSF patterns. The three different realisations all have stellar density 
 $\sim 1/{\rm arcmin}^{2}$. The maps in the first row show the ``true'' PSF ellipticity 
 ($\epsilon_{1}$) field that we would like to reconstruct from the stellar data in the 
 second row, the observed stellar ellipticities. The last three rows show model  PSF 
 ellipticity maps constructed with \psfent, a 5th-order polynomial fit and a 5$\times$
 5-pixel boxcar smoothing, respectively.}
 \label{fig:psfpattern_map}
\end{figure*}

\begin{figure}
\begin{center}
   \subfigure[]{\includegraphics[scale=0.35]{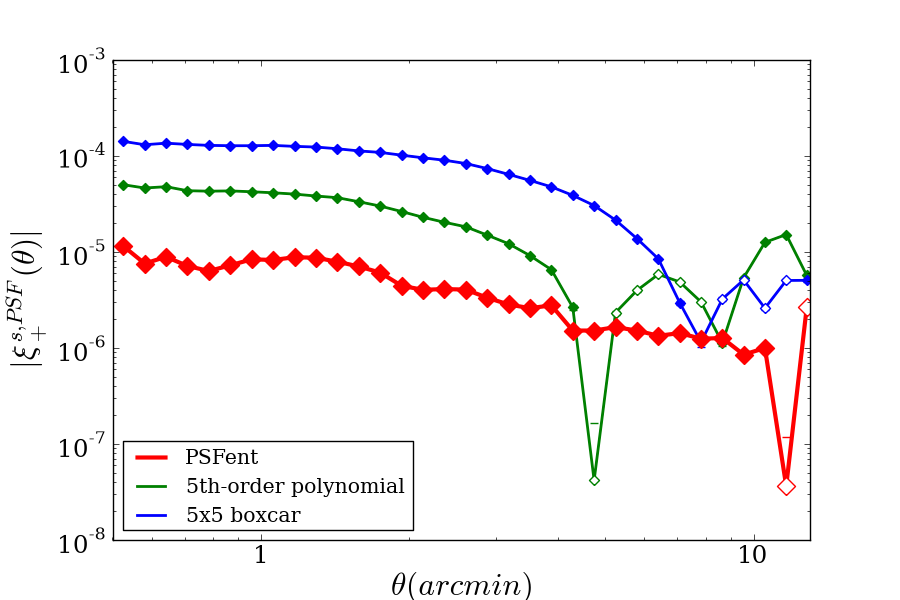} } \\
   \subfigure[]{\includegraphics[scale=0.35]{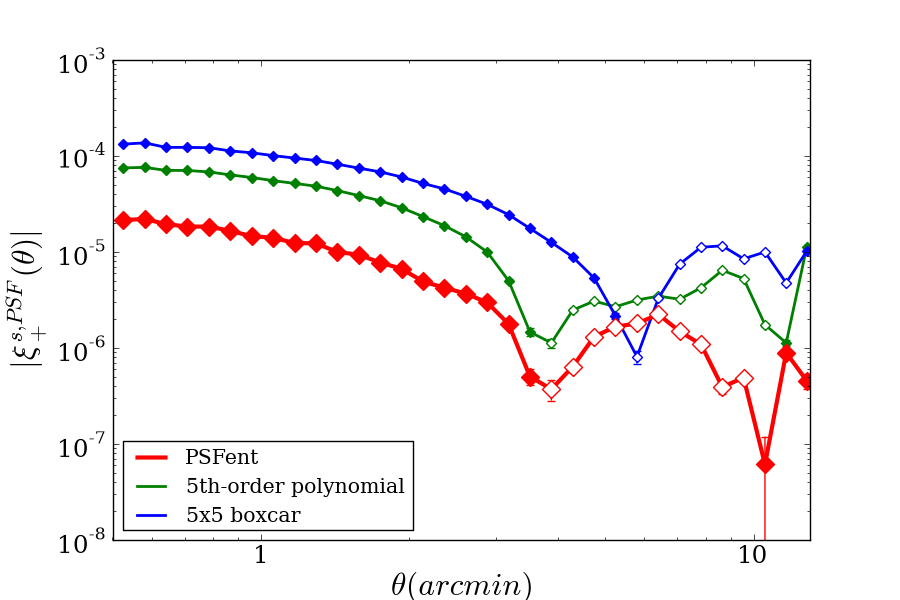} } \\
   \subfigure[]{\includegraphics[scale=0.35]{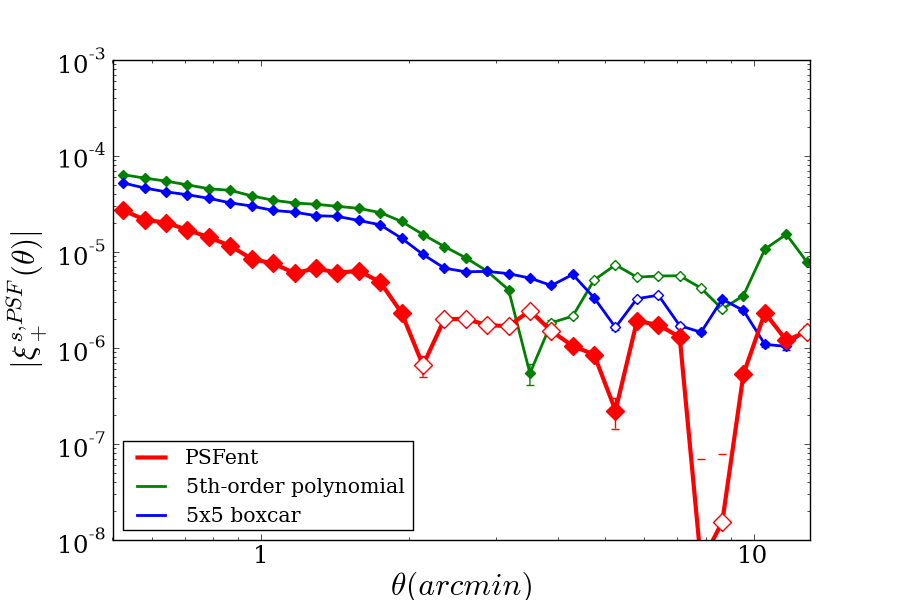} }
 \caption{Absolute two-point correlation functions of the PSF ellipticity model errors
 for the three different PSF patterns in \Fref{fig:psfpattern_map}. In each panel, we 
 show results for \psfent (red),  5th-order polynomial fitting (green) and 5$\times$5-pixel 
 boxcar smoothing (blue). The hollowed labels indicate negative values.}
 \label{fig:psfpattern_corr}
\end{center}
\end{figure}

The 100 atmospheric realisations in our simulation suite provide a wide range 
of PSF patterns. For example, three very different PSF patterns on a single 
LSST CCD sensor are shown in the top row of \Fref{fig:psfpattern_map}, 
where the colours represent true PSF $\varepsilon_{1}$ values. 
We treat the $\varepsilon_{1}$ and $\varepsilon_{2}$ patterns as they were 
independent, as noted earlier, and only show $\varepsilon_{1}$ here.  

We observe that the left column (a) contains a PSF pattern with some large-scale 
stripes that vary smoothly across the CCD sensor, with very fine ripples 
aligned at a direction different from the large stripes; the middle column (b) 
contains medium-size blobs without any preferred direction; finally, the right 
column (c) shows nearly equal strength of stripes in two nearly orthogonal 
directions, creating a grainy high-frequency pattern. When attempting to model 
these patterns at a typical galactic latitude of $|b|\sim60$ (stellar density 
$\simeq 1 /{\rm arcmin}^{2}$), the stars available to us for reconstructing these 
PSF patterns are shown in the second row on the same ellipticity colour scale. 
The last three rows of \Fref{fig:psfpattern_map} show the PSF model generated 
from three different PSF interpolation techniques in the following order: \psfent, 
5th-order polynomial fitting and $5\times 5$-pixel boxcar smoothing.

Visually, one can readily see the power of \psfent in modelling the very different 
PSF patterns over the other two methods. In (a), all three methods failed to 
model the fine ripples, since they are much finer than the average stellar 
separations. They all do, however, manage to pick up the smooth ``stripe'' 
component. Both the polynomial and boxcar model in this case show bad 
behaviours on the edges of the field, due to the small number of ill-measured stars 
dominating the model. In (b), where the underlying PSF patterns consist of mainly 
medium scale ``patches'', the Gaussian ICF used in \psfent enables the model to 
capture these features nicely, which is not possible with a polynomial model. The 
boxcar model, on the other hand, is limited by the size of the filter, which in this 
case is slightly larger than the patches in the patterns. Finally in the right column 
(c), we show an example where the polynomial model becomes worse than even 
a simple boxcar smoothing. In this case the PSF pattern almost has no power on 
the large scales, causing the polynomial model to be entirely dominated by the 
noise. 

In all the cases shown here, we can see the flexible multi-scale algorithm allows 
\psfent to model structures on a large number of scales, and is well regulated by 
the prior construction thus less sensitive to noise.

The absolute correlation functions $|\xi_{+}^{s,\rm PSF}|$ for the three examples 
in \Fref{fig:psfpattern_map} are shown in \Fref{fig:psfpattern_corr}. In general, the 
model errors are more correlated on small scales due to the sparse sampling, and 
the limited resolution for all three modelling techniques. The shape of these 
correlation functions can be either smooth or oscillating. In particular, for polynomial 
models, the shape of $\xi_{+}^{s,\rm PSF}$ has some characteristic features: a 
transition from positive (correlation) to negative (anti-correlation) always appear 
at 3$'$ -- $4'$. As discussed in H12, this is a result of both the modelling method and 
the true atmospheric PSF pattern. 

\begin{figure}
   \subfigure{\includegraphics[scale=0.35]{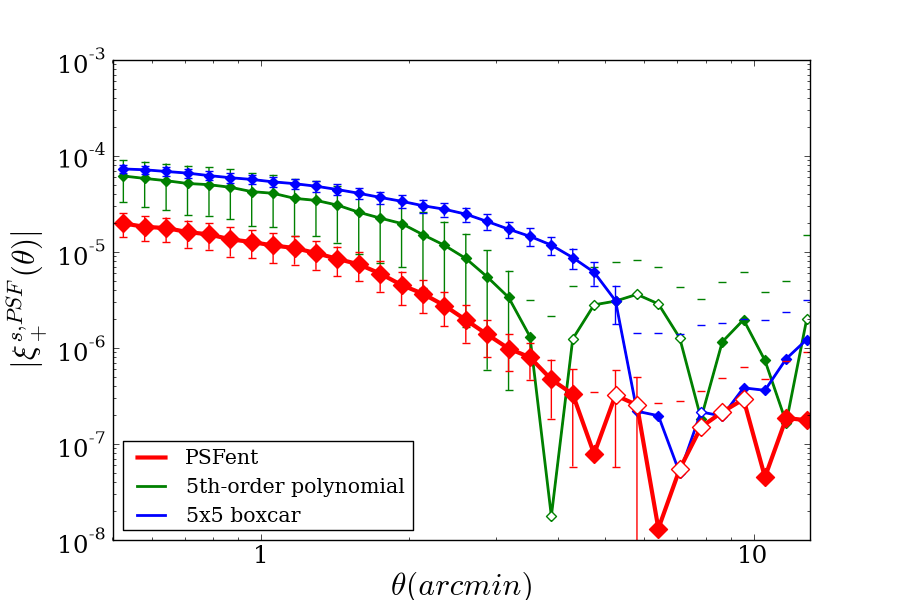} }
 \caption{Median correlation function of the PSF ellipticity model errors of
 100 different PSF patterns, at the ``typical'' stellar density of $1/{\rm
 arcmin}^{2}$. The results are shown for \psfent (red), 5th-order polynomial
 fitting (green) and 5$\times$5 boxcar filtering (blue). The error bars indicate 
 the rms spread in the 100 exposures divided by $\sqrt{100}$. Hollowed labels 
 indicate negative values.}
 \label{fig:psfpattern_corr_average}
\end{figure}

\Fref{fig:psfpattern_corr_average} shows, for the 100 different realisations of the 
atmosphere, the median behaviour of these ellipticity error correlation functions 
with the error bars indicating the standard deviation of the 100 curves divided by 
$\sqrt{100}$. The median $\sigmaePSF$ and $\sigmasyssq$ values for these 100 
atmospheric realisations are listed in \Tref{tab:stats_psfpattern_average}.
The final column $F_{\rm sys}$, as explained earlier, is a measure of 
the {\it relative} reduction in spatially-correlated residual ellipticity for the level of 
spatial correlation in the model errors independent of the absolute errors. 
Quantitatively, \psfent provides $\sim17\%$ improvement in the absolute ellipticity 
modelling error, or $\sigmaePSF$, over the 5th-order polynomial and $\sim22\%$ 
improvement over the boxcar smoothing model. For the correlation of these errors, 
or $\sigmasyssq$, \psfent performs $\sim3.5$ times better than the polynomial 
model and $\sim7$ times better than boxcar smoothing. The corresponding $F_{\rm sys}$ 
values suggest that the model errors from \psfent is $\sim2.5$ times less correlated 
than that from polynomial models and $\sim4.2$ times less correlated than that from 
boxcar models. Notice that the main power in \psfent lies not in the absolute 
reduction of the model errors, but in the fact that the flexible free-form model creates 
makes errors less correlated in space, which is an important property for 
measurements like cosmic shear, where the main signal is embedded in the spatial 
correlation of galaxy shapes.    

\begin{table}
  \centering
  \begin{tabular}{c | c | c | c}
               &$\sigmaePSF$ 
               &$\sigmasyssq$ 
               & $F_{\rm sys}$  \\ \hline                           
    \psfent          & 7.41$\times 10^{-3}$   &2.41$\times 10^{-6}$
    & 4.39$\times 10^{-2}$ \\ 
    Polynomial  &  8.93  $\times 10^{-3}$ & 8.67$\times 10^{-6}$ 
    & 10.87$\times 10^{-2}$ \\ 
    Boxcar          &   9.51 $\times 10^{-3}$ & 16.77$\times 10^{-6}$ 
    & 18.54$\times 10^{-2}$\\ \hline 
  \end{tabular}
  \caption{Median metric $\sigmaePSF$ and $\sigmasyssq$ for the three PSF interpolation 
  techniques for 100 different PSF patterns sampled at the nominal stellar density of 
  $1/{\rm arcmin}^{2}$. The final column $F_{\rm sys}$ is a measure of the level of the spatial 
  correlations in the PSF model errors, independent of the absolute errors.}
  \label{tab:stats_psfpattern_average}  
\end{table}


\subsection{Variation with stellar density}
\label{sec:Results_density}

\begin{figure*}
  \begin{center}
   \includegraphics[height=1.6in]{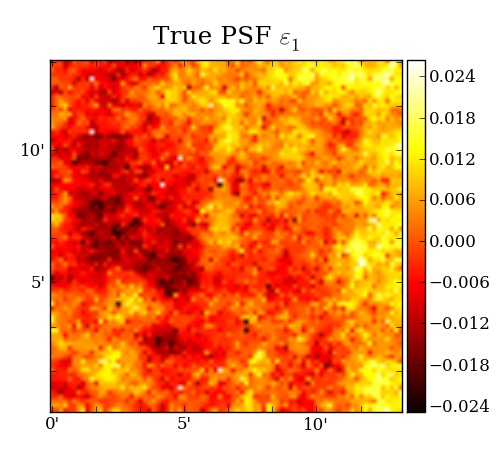} 
   \includegraphics[height=1.6in]{figs/target_003_1_e1.png}     
   \includegraphics[height=1.6in]{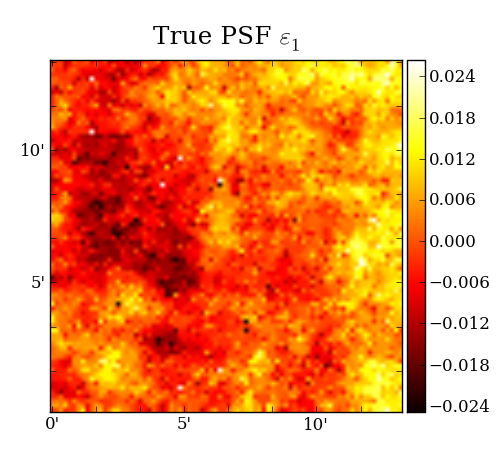}     \\ 
   \includegraphics[height=1.6in]{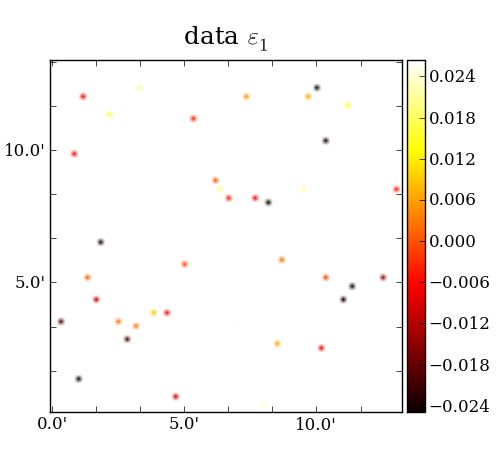}  
   \includegraphics[height=1.6in]{figs/star_003_1_e1.png}  
   \includegraphics[height=1.6in]{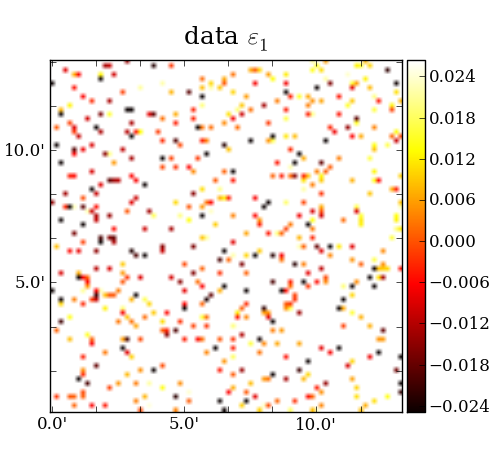}    \\ 
   \includegraphics[height=1.6in]{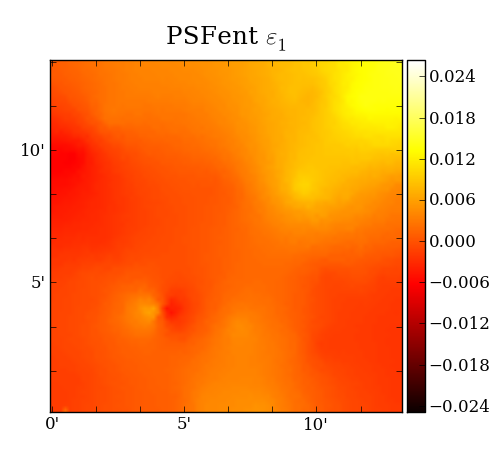} 
   \includegraphics[height=1.6in]{figs/target_003_1_model_psfent_e1.png}   
   \includegraphics[height=1.6in]{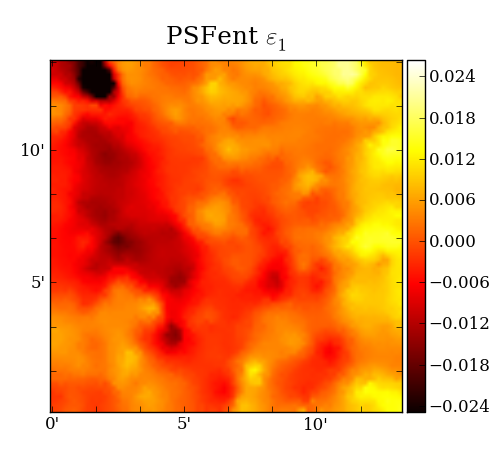}   \\
   \includegraphics[height=1.6in]{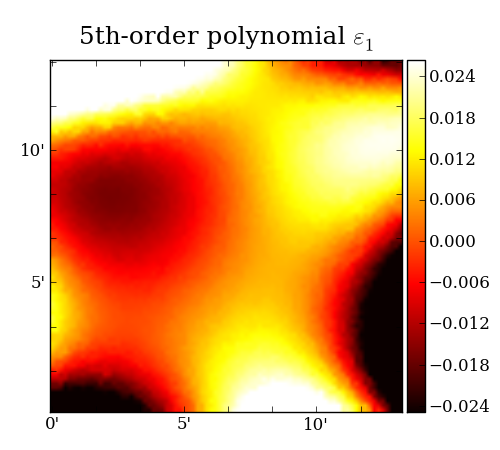}   
   \includegraphics[height=1.6in]{figs/target_003_1_model_p5_e1.png}   
   \includegraphics[height=1.6in]{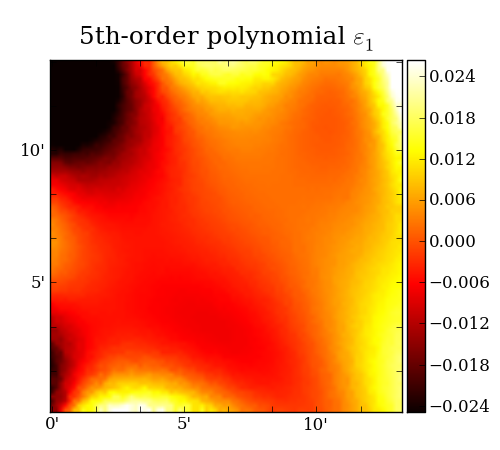}   \\ 
   \subfigure[]{\includegraphics[height=1.6in]{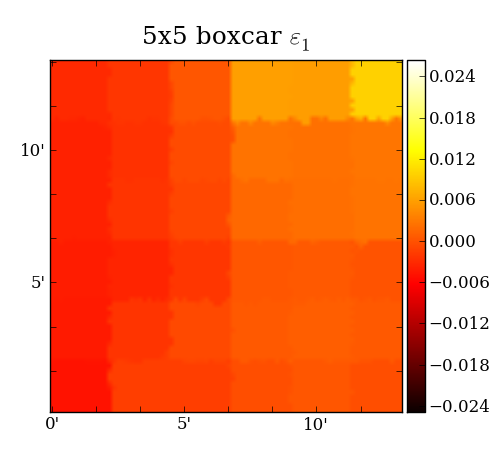} }  
   \subfigure[]{\includegraphics[height=1.6in]{figs/target_003_1_model_b5_e1.png} }  
   \subfigure[]{\includegraphics[height=1.6in]{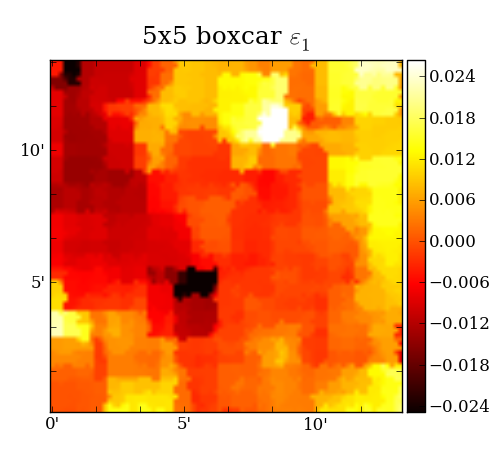} }  \\
 \end{center}
 \caption{Illustration of the single short exposure PSF interpolation problem,
 and the performance of different interpolation methods as a function of
 stellar density. See the caption of \Fref{fig:psfpattern_map} for the
 description of the maps in the five rows; the different columns correspond to the same
 atmospheric PSF pattern (\Fref{fig:psfpattern_map}(b)), but sampled by stars with 
 densities of 0.25 (a), 1 (b) and 4 (c) ${\rm /arcmin}^{2}$, as shown in the 
 second row.}
 \label{fig:psfpattern_map2}
\end{figure*}

\begin{figure}
 \begin{center}
   \subfigure[]{\includegraphics[scale=0.35]{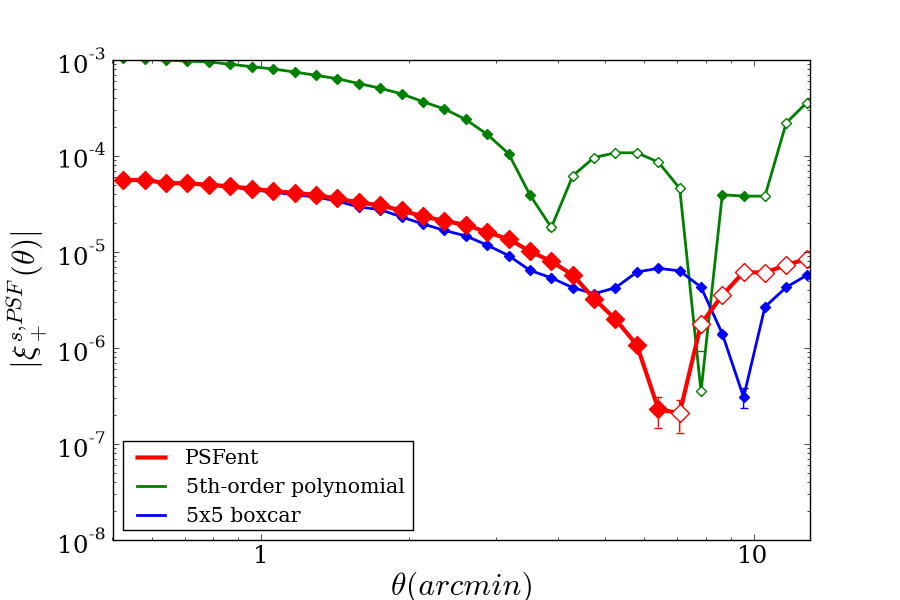}} \\
   \subfigure[]{\includegraphics[scale=0.35]{figs/corr_003_1.png} } \\
   \subfigure[]{\includegraphics[scale=0.35]{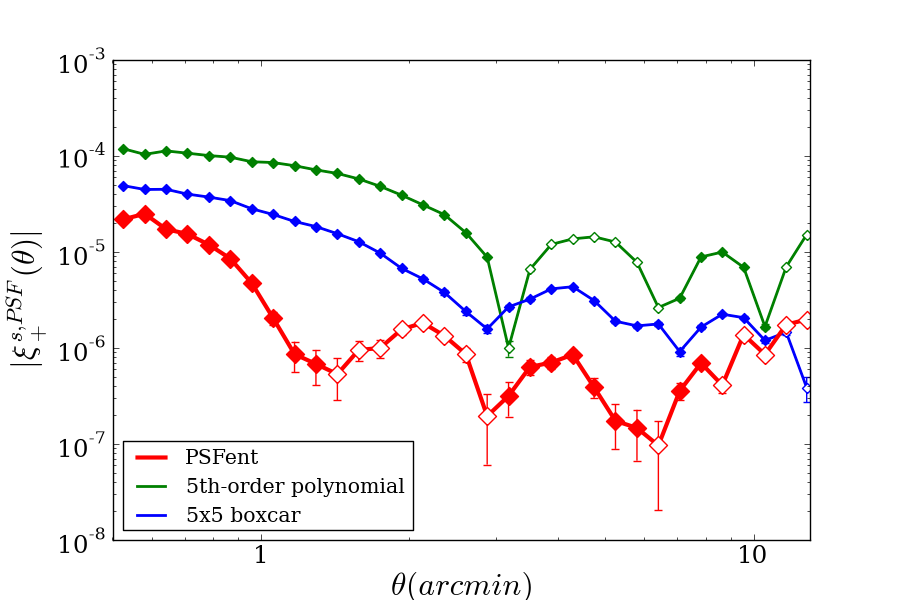} }     
 \end{center}
 \caption{Absolute two point correlation functions of the PSF ellipticity model errors 
 for different stellar densities: (a) 0.25, (b) 1 and (c) 4 ${\rm /arcmin}^{2}$. The three 
 cases have the same underlying PSF pattern but are sampled at different rates. In 
 each panel, we show results for \psfent (red), 5th-order polynomial fitting green) and 
 5$\times$5 boxcar smoothing (blue). The hollowed labels indicate negative values.}
 \label{fig:psfpattern_corr2}
\end{figure}

\begin{figure*}
\begin{center}
\subfigure[]{\includegraphics[scale=0.35]{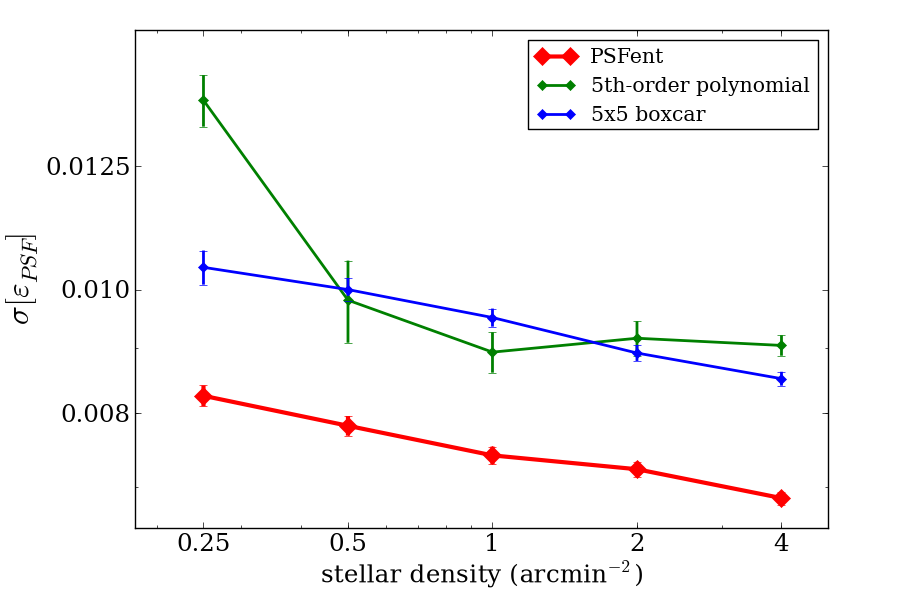} } 
\subfigure[]{\includegraphics[scale=0.35]{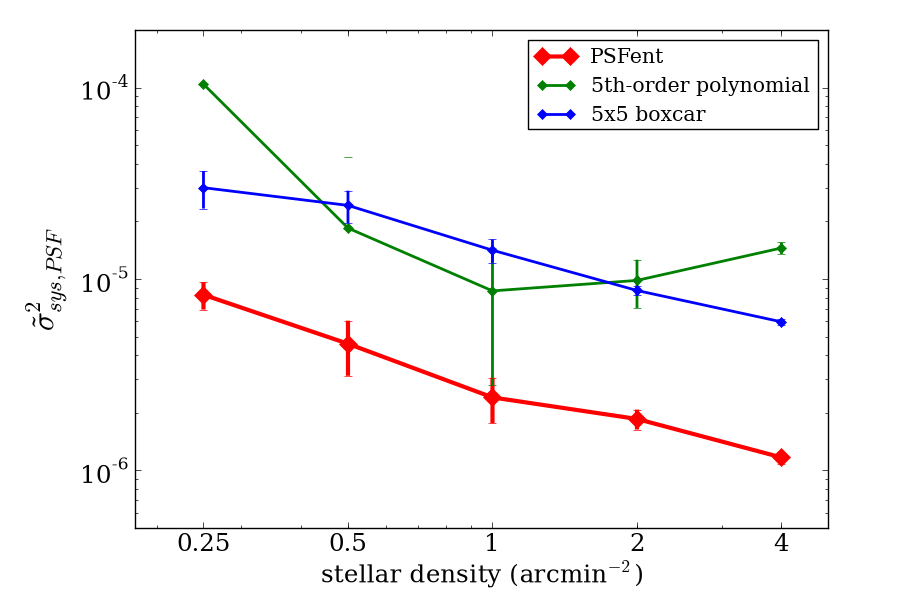} }
\caption{Median values for 100 PSF patterns for the two statistics:
(a) $\sigmaePSF$ and (b) $\sigmasyssq$ over a range of stellar densities. 
The error bars show the interquartile range divided by $\sqrt{100}$. We show in 
 each panel results for \psfent (red), 5th-order polynomial fitting (green) and 
 5$\times$5 boxcar smoothing (blue). In both statistics, \psfent performs consistently 
 better than the other two techniques.}
 \label{fig:stats}
\end{center}
\end{figure*}

Having examined the performance of the three PSF interpolation methods  on a
``typical'' field, we would now like to understand how the three PSF 
interpolation schemes are affected by the available density of stars (we explore 
the range from 0.25 to 4 stars per arcmin$^{2}$ for a complete sample of 
realistic stellar distributions). This test is especially important for images at high 
galactic latitude, where stars are very sparse. The ability to reconstruct the PSF 
variation at these fields may increase the effective area of a survey and  therefore 
its statistical power. \Fref{fig:psfpattern_map2} shows, for the PSF pattern in 
\Fref{fig:psfpattern_map}(b), how the PSF model improves, for the three interpolation 
methods, as the stellar density increases. \Fref{fig:psfpattern_corr2} shows how the 
residual ellipticity correlation in each case changes accordingly. 

\Fref{fig:psfpattern_map2} visually illustrates one example of how the three different 
interpolation methods respond to the increased available stellar data points. We 
observe that the polynomial models appear to be particularly ill behaved when the 
available stars are under-dense (a) and over-dense (c). This is an example of 
imposing an improper prior assumption about the PSF pattern while ignoring the data. 
In contrast, the simple boxcar smoothing technique works in the opposite direction, 
where the model is purely driven by data with essentially no assumption on the 
expected PSF patterns. As a result, the models are just reflecting the available data, 
where we get a model with no structure in the under-dense case (a), and a model with 
lots of small scale structure in the over-dense case (c). \psfent is effectively a more 
sophisticated version of the boxcar smoothing with informative priors and multiple 
structure scales. The change from (a) to (c) for the \psfent model is qualitatively similar 
to the boxcar smoothing, as it is primarily dictated by data. But when data is 
insufficient, as in (a), \psfent does not generate entirely flat models like boxcar 
smoothing, rather, we can see traces of the \psfent priors creating some structure in 
the PSF model. When the stellar data is abundant (c), \psfent lets data take over to 
drive the fit and only makes sure that the model stays in a physically reasonable 
range as specified by the priors. \Fref{fig:psfpattern_corr2} confirms the above 
observation more quantitatively.

In \Fref{fig:stats}, we show for our 100 different atmosphere realisations the 
median $\sigmaePSF$ and $\sigmasyssq$ statistics as a function of stellar 
density. In all stellar densities, \psfent consistently performs $\sim20\%$ better 
in $\sigmaePSF$ and 3--10 times better in $\sigmasyssq$ compared to boxcar 
smoothing and polynomial fitting. 

The two statistics show similar trends in general. As we explained earlier, for 
\psfent and boxcar smoothing, since the model is primarily driven by data, the 
model improves monotonically as more data becomes available. The 20\% 
and nearly 4-times improvement of \psfent in $\sigmaePSF$ and $\sigmasyssq$ 
respectively compared to boxcar smoothing mainly comes from \psfent's ability to 
capture multi-scale structures and regulate the model using priors so that noise 
does not get amplified. A fixed order polynomial function, on the other hand, when 
optimised for a certain stellar density (1$/ \rm arcmin^{2}$), over-fits 
(\Fref{fig:psfpattern_map2}(c)) or under-fits (\Fref{fig:psfpattern_map2}(a)) data when 
the stellar density varies, which results in a local minimum in the two 
green curves. The polynomial model, when optimised, is a reasonably good 
description of the smooth variation in the large-scale PSF patterns, but still fails to 
capture the small-scale structures, which explains why \psfent still performs better 
in that case. 

\chihway{Although we emphasise the improvement of \psfent over the other two methods, 
it is important to note at this point that the main improvement here is not from the specific 
type of functional form (or un-parametrised model) one uses, but rather, the use of either 
realistic simulations or calibration data to inform the prior PDFs for the flexible model 
parameters. We chose pixelated maps for use in \psfent because we learned from 
simulations that the PSF patterns are complicated  and requiring a very flexible model.
One can imagine, for example, an alternative method with the same spirit, where a basis 
of high-order polynomials are used to reconstruct the PSF variations with coefficients 
constrained by priors derived from simulations.} 



\section{Discussion}
\label{sec:FurtherWork}



\subsection{Implications of $\sigmaePSF$ and $\sigmasyssq$ on shear 
systematics}
\label{sec:Result_ShearSys}
To project the improvement on PSF interpolation from \psfent onto the 
improvement in weak lensing shear measurements for a LSST-like survey, we 
return to \Tref{tab:stats_psfpattern_average} and discuss the implications of 
the $\sigmaePSF$ and $\sigmasyssq$ values under the nominal stellar 
density of 1$/{\rm arcmin}^{2}$. 

\chihway{According to \citet[][hereafter AR08]{AR08}, for future stage IV ground-based 
weak lensing surveys\footnote{\chihway{In \citet{2006astro.ph..9591A}, LSST is classified 
as a Stage IV ``Large Survey Telescope (LST)'' project. The Stage IV LST model assumes 
a ground-based survey with half-sky coverage, median readshift 1.0 -- 1.2, and 30 -- 40 
well-measured weak lensing galaxies per arcmin$^{2}$.}} \citep{2006astro.ph..9591A} 
not to be systematics-limited, one 
can set limits on the allowed systematic errors on the spurious shear power spectrum.} 
\cite{2008A&A...484...67P} extended from AR08 and estimated that the allowed 
errors on determining the PSF ellipticity corresponding to those limits on spurious shear 
power spectrum is:
\begin{equation}
\sigmaePSF \leq 10^{-3}
\label{eq:limits1}
\end{equation}

Combining the first column in \Tref{tab:stats_psfpattern_average} and \Eref{eq:scale1}, 
we can derive the number of exposures needed for each of the interpolation methods to 
achieve \Eref{eq:limits1} as listed in \Tref{tab:Nexp}. Also listed is the corresponding 
operation time for LSST, where we have adapted the assumptions in C12 and assumed 
two different scenarios -- ``optimistic'' ($N_{\rm exp}=368$) and ``pessimistic'' 
($N_{\rm exp}=184$), where a total of $N_{\rm exp}$ single 15-second exposures are 
combined in the final 10-year dataset for cosmic shear measurements. 

\begin{table}
  \centering
  \begin{tabular}{c | c | cc}
               &$N_{\rm exp}$ &\multicolumn{2}{c}{operation time (years)}  \\                          
               & & optimistic & pessimistic   \\     \hline
    \psfent          & 55 & 1.50 & 2.99 \\ 
    Polynomial  & 80 & 2.17 & 4.35 \\ 
    Boxcar          & 90 & 2.45 & 4.89 \\ \hline 
  \end{tabular}
  \caption{Number of exposures required for the PSF ellipticity measurement accuracy 
  to meet \Eref{eq:limits1} for the three PSF interpolation methods under nominal conditions. 
  Also listed are the corresponding expected time span these exposures can be obtained by 
  LSST in the optimistic and pessimistic scenarios.}
  \label{tab:Nexp}  
\end{table}
\begin{table}
  \centering
  \begin{tabular}{c | c | cc}
               &$N_{\rm exp}$ &\multicolumn{2}{c}{operation time (years)}  \\                          
               & & optimistic & pessimistic   \\     \hline
    \psfent          & 105 & 2.86 & 5.72 \\ 
    Polynomial  & 368 & 10.0 & 20.0 \\ 
    Boxcar          & 710 & 19.34 & 38.7 \\ \hline 
  \end{tabular}
  \caption{\chihway{Number of exposures required for the PSF ellipticity measurement 
  accuracy to meet the target value set by AR08 and C12 for the three PSF interpolation 
  methods under nominal conditions. Also listed are the corresponding expected time 
  span these exposures can be obtained by LSST in the optimistic and pessimistic 
  scenarios.}}
  \label{tab:Nexp2}  
\end{table}

However, we note that \Eref{eq:limits1} is a rather simplistic estimation that does 
not account properly for the correlation properties of these errors and the spurious 
shear arising from a realistic PSF correction pipeline. An alternative and more 
realistic approach to interpret the results from our study in terms of shear 
measurements is to turn to C12, where we looked at the additive spurious shear 
correlation function by actually measuring these levels on high fidelity simulations. We 
concluded in C12 that the polynomial PSF model generates spurious shear correlation 
approximately at the level required by AR08 in the optimistic case and 2 times 
too high in the pessimistic case. 
\chihway{Since \psfent provides a $\sim3.5$ times improvement in the PSF error 
correlation over polynomial models (\Tref{tab:stats_psfpattern_average} second column), 
we can expect it to also lower the spurious shear power spectrum by a factor of $\sim3.5$ if 
\textit{all} the spurious shear is due to PSF interpolation errors. This brings the level of 
spurious shear power spectrum 3.5 (optimistic) and 1.75 (pessimistic) times lower than the 
target level. Combined with \Eref{eq:scale2}, we summarise in \Tref{tab:Nexp2} the results 
in terms of the number of exposures and survey time needed to achieve the target value set 
in AR08. }

\chihway{
On the other hand, if one considers that there are shear measurement algorithm errors 
that have not been accounted for in C12, then the situation could be worse. Assume, for 
example, these unaccounted algorithm errors take up to $50\%$ the allowed systematic 
errors set by AR08, the improvement in \psfent then results in a spurious shear 
power spectrum about 1.3 times lower (optimistic) and 1.1 times higher (pessimistic) than 
the target level. In that case, even \psfent will be a marginal failure in the pessimistic 
scenario.}

\subsection{Correlating galaxies across exposures}

Correlating galaxy ellipticities in different exposures has been one of the proposed 
solutions to the problem of stochastic atmospheric PSF correlations 
\citep{2006JCAP...02..001J}. This, however, comes at the price of decreasing the 
statistical power of the survey. Exploring the tradeoff between statistical and 
systematic PSF interpolation errors using this technique is not the main focus of this 
paper. Here, we have effectively assumed a \lensfit-style 
\citep{2007MNRAS.382..315M} analysis, where the PSF ellipticity map is estimated 
for each exposure, the galaxy images deconvolved, and the ellipticity estimates simply 
averaged. This assumption is justified by the behaviour of the residual PSF ellipticity 
correlation function with increasing $N_{\rm exp}$ as we have shown. We presume that 
when correlating between different exposures, it will still be beneficial to start with a 
more accurately-interpolated PSF model.


\subsection{Computational cost}

On a standard 64-bit 2$\times$2.2~GHz processor, \psfent takes on 
average $\sim$13.0 seconds per LSST CCD sensor per pair of shape parameters
($\varepsilon_{1},\varepsilon_{2}$). Adding error estimation (sampling from the 
posterior cloud) adds an extra $\sim$9 sec of run time. This is about an order
of magnitude slower than the two other techniques we investigated. For more
complicated shape parametrisation the runtime would increase linearly with the
number of parameters, and also with the number of images analysed. For
example, if \psfent were to be used for in the LSST data processing pipeline,
the PSF model in each exposure on the $\sim$200 CCD sensors would need to be
calculated in $\sim$ 19 seconds (15-second exposure + 2-second readout + 
2-second shutter open/close). This demands
$\sim$200 computers running for the PSF reconstruction alone. Though large,
these numbers are not outrageous considering the expected decrease in unit cost 
for computers over the next decade and the possibility of further accelerating the 
code via hardware parallelisation such as GPUs. 

In the meantime, we recommend that the current code can be used for smaller
scale  datasets when extremely accurately interpolated PSF maps are vital for
the specific science goal. As well as weak lensing, these might include 
reconstructing the detailed structures of complex objects, and precision
photometry of faint objects. 


\section{Conclusions}
\label{sec:Conclusions}
In this paper we have introduced a new PSF interpolation method, \psfent, based 
on a multi-scale maximum-entropy image reconstruction code. The problem  we set
out to solve is reconstructing the multi-scale spatial variations of PSF 
shapes, due to atmospheric turbulence, in short exposure images, from sparsely
distributed, noisily measured stars -- a potential problem for future weak
lensing surveys  such as LSST.

Our analysis of simulated data allows us to draw the following conclusions:

\begin{itemize}


  \item Compared with two other PSF interpolation methods, one of which 
  is commonly used in current data analyses, \psfent 
  provides more accurately interpolated PSF shapes: the absolute residual 
  ellipticities improve $\sim20\%$ while the correlated residual ellipticities are 
  a factor 3 --10 smaller than the other two methods, over a wide range of different 
  PSF patterns and stellar densities.
  \item When combining multi-epoch datasets, the interpolation errors due to
  the atmosphere are stochastic, decreasing with the number of exposures
  taken as $1/\sqrt{N_{\rm exp}}$. The correlation function amplitude decreases 
  as $1/N_{\rm exp}$.
  \item \chihway{The improvement in PSF modelling from \psfent suppresses the 
  spurious shear in cosmic shear measurements. Combining previous studies from 
  realistic simulations and the results in this work, we estimate that systematic errors 
  due to PSF interpolation for LSST will be 3.5 (optimistic) and 1.75 (pessimistic) times 
  lower than the statistical errors.}
  \item \chihway{Taking into account other algorithm errors, however, 
  the improvement from \psfent may not be sufficient to bring the systematic errors 
  down to the target level in the most pessimistic scenario.}  
\end{itemize}

While it may still become a practical solution even for large datasets such as
LSST, the relatively high computational cost of powerful algorithms like
\psfent should motivate the development of faster inference methods. The
paper by \citet{2012MNRAS.419.2356B} appeared as we were completing this 
study: their results are complementary to those presented here, in that they 
investigate the PSF interpolation problem on larger scales, using re-sampled real 
Subaru data that is less dominated by the atmosphere. It would be very interesting 
to apply their Kriging technique to our simulated data, and compare performance
in terms of accuracy and speed. Like \citet{2012MNRAS.419.2356B}, we have 
demonstrated the benefits of using very flexible models, but also the use of methods 
both inspired and constrained by realistic simulations. Injecting statistical information 
about the atmospheric PSF anisotropy into interpolation methods appears to be a 
fruitful approach.
\vspace{0.1in}

All the simulated data used in this paper is freely available at the following website: 
\url{http://www.slac.stanford.edu/~chihway/psfent/}.


\section*{Acknowledgments}

LSST project activities are supported in part by the National Science
Foundation through Governing Cooperative Agreement 0809409 managed by the
Association of Universities for Research in Astronomy (AURA), and the
Department of Energy under contract DE-AC02-76-SFO0515 with the SLAC
National Accelerator Laboratory. Additional LSST funding comes from private
donations, grants to universities, and in-kind support from LSSTC Institutional
Members.

PJM acknowledges support from the Royal Society in the form of a university
research fellowship.
This work was supported in part by the U.S.\ Department of Energy under
contract number DE-AC02-76SF00515. 

We thank Catherine Heymans, Barney Rowe and Lance Miller for useful 
discussions; and Catherine Heymans and Barney Rowe for making their CFHT 
results available before publication. We also thank Beth Willman, Robert Lupton 
and David Wittman for useful comments which have helped improve this paper 
substantially.
                
\label{lastpage}


\bsp

\appendix

\section{Other modelling methods} 
\label{sec:methods}


\subsection{Polynomial fitting}

As discussed in \Sref{sec:Background}, in most current weak lensing 
pipelines, the PSF spatial variation is assumed to be smoothly varying 
on scaled comparable to the field and can thus be modelled with some 
low order polynomial functions. We have shown that PSF patterns from 
short exposures, however, display higher frequency spatial variations that 
require higher order fits. As a result, we have examined polynomials of 
order 2 to 5 as PSF models and found that 2nd-order models are insufficient 
in representing the PSF variations; 3rd-order models are usually 
sufficient, but occasionally an even higher order model is needed. 
We choose to use 5th-order models in our main analyses, but have 
confirmed that in most cases, the results are identical to 3rd-order fits.
%
We minimise the effective $\chi^{2}$:
\begin{equation}
 \chi^{2}=\sum w_{\varepsilon}^{2}(\varepsilon^{model.j}_{i}
 (\boldsymbol{p};x,y)-\varepsilon_{i})^{2}, 
 \; i=1,2 \;,
\end{equation}
where $\varepsilon^{model,j}_{i}(\boldsymbol{p};x,y)$ is a two dimensional, 
$j$th order polynomial function of $(x,y)$, $\boldsymbol{p}$ are the 
fitting parameters and we use the signal-to-noise ratio of each star as 
$w_{\varepsilon}$.

%
%


\subsection{Boxcar smoothing}
 
Alternatively, we examine the approach of modelling the PSF by directly 
smoothing the stellar ellipticities with a boxcar filter. In this approach, 
stars need to first be binned into large pixels, where we assign a weighted 
average ellipticity to each pixel $(j_{1},j_{2})$, defined:
\begin{equation*}
 \overline{\varepsilon^{j_{1}j_{2}}_{i}}=\sum_{pixel\;(j_{1},j_{2})} W_{\varepsilon}
 \varepsilon_{i}\;,\;\; i=1,2 \;.
\end{equation*}
The size of the pixels are determined by the number of stars -- we pixilate the 
image so that in each pixel contains approximately 1 star. A boxcar filter of size 
$m \times m$ pixels is then applied to the large pixel grid so that the 
ellipticity of pixel $(j_{1},j_{2})$ is replaced by the average ellipticity of the 
neighbouring $(m^{2}-1)$ pixels:
\begin{align*}
 \varepsilon^{model,j_{1}j_{2}}_{i}=&\frac{1}{m^{2}-1}((\sum_{a=-m'}^{m'} 
 \sum_{b=-m'}^{m'} \overline{\varepsilon^
{(j_{1}+a)(j_{2}+b)}_{i}}) -\overline{\varepsilon^{j_{1}j_{2}}_{i}})\;;\\
 m'=&\frac{m-1}{2},\; i=1,2\;.
\end{align*}
The original coordinates are then recovered so that the modelled PSF ellipticities of 
all points falling in pixel $(j_{1},j_{2})$ will be $\varepsilon^{model,j_{1}j_{2}}$ 


\section{PSFENT with other shape parameters}
\label{sec:R_test}
\chihway{
It had been shown in H12 that due to atmospheric distortion, the size of the PSF varies 
on spatial frequencies similar to that of $\varepsilon_{1}$ and $\varepsilon_{2}$. This 
suggests that with the same procedure described in \Sref{sec:Programme}, we could 
use \psfent to interpolate the PSF size, and by extension other shape parameters. We 
demonstrate below our results of interpolating the shape parameter ``FWHM\_WORLD'' 
output from Source Extractor (which is an estimate of the PSF FWHM size in arcseconds) 
with the three interpolation methods used in \Sref{sec:Results}. }

\begin{figure}
  \begin{center}
   \includegraphics[height=1.6in]{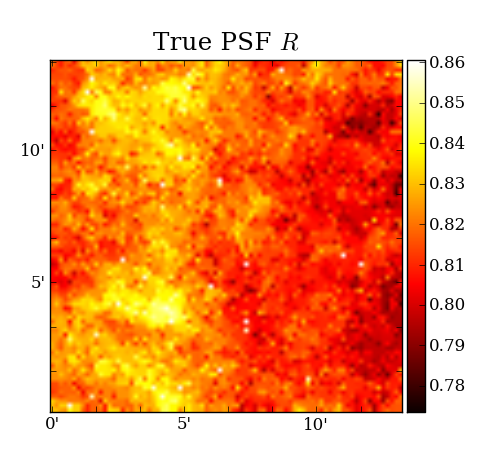} \\
   \includegraphics[height=1.6in]{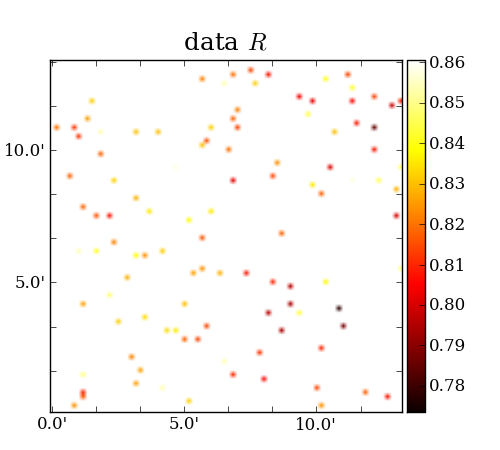}   \\ 
   \includegraphics[height=1.6in]{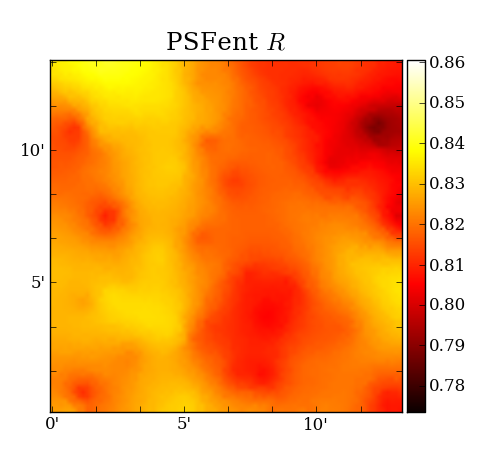} \\
   \includegraphics[height=1.6in]{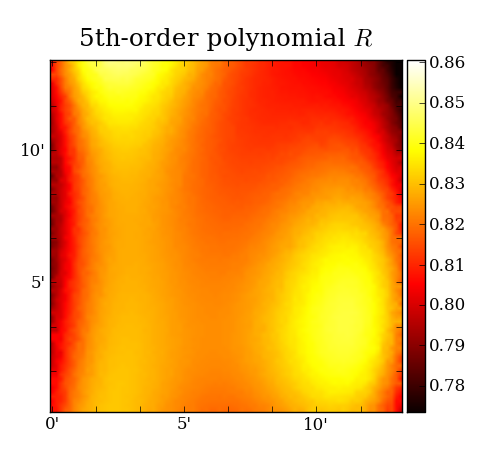}\\ 
   \includegraphics[height=1.6in]{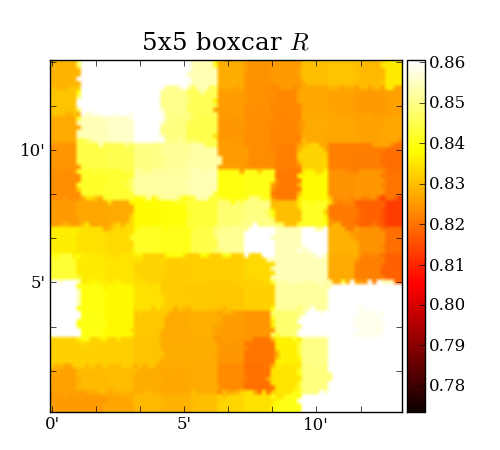} \\
 \end{center}
 \caption{\chihway{Illustration for interpolation of PSF size (FWHM size in arcseconds). The top map 
 shows the ``true'' PSF size field that we would like to reconstruct from the stellar data in 
 the second map, the observed stellar size. The next three maps show model maps 
 constructed with \psfent, a 5th-order polynomial fit and a 5$\times$ 5-pixel boxcar smoothing, 
 respectively. The atmospheric PSF pattern and stellar density is the same as that in 
 \Fref{fig:psfpattern_map}(b).}}
 \label{fig:psfpattern_map3}
\end{figure}

\chihway{First we derive the appropriate priors ($m_{i}$ values) for FWHM\_WORLD from 
the ``true'' PSF maps. We have $m_{i}$=[0.0001, 0.0002, 0.0026, 0.0055, 0.0079, 0.0100, 
0.0206] for FWHM\_WORLD. The set of priors for the FWHM\_WORLD variation appear to 
follow the same trend as that of $\varepsilon_{1}$ and $\varepsilon_{2}$, while putting 
slightly more power on the larger scales. This can also be observed qualitatively in 
\Fref{fig:psfpattern_map3}, where we show the equivalent of \Fref{fig:psfpattern_map}(b), 
with the interpolant being the PSF size, $R$ (or FWHM\_WORLD), instead of $\varepsilon_{1}$. 
We then calculate for the 100 atmosphere realisations and for the three different PSF 
interpolation methods, the statistics in analogy with $\sigmaePSF$:} 
\begin{equation}
   \Delta=\frac{\sigma[\delta R^{2}]}{\langle R^{2}\rangle}
  =\frac{\sqrt{\langle (\delta R^2)^{2}\rangle}}{\langle R^{2}\rangle},
\end{equation}
\noindent where 
\begin{equation}   
  \delta R^2= |R^{2}_{model}- R^{2}_{true}| \; ;
\end{equation}  
\chihway{
We find that the median $\Delta$ values for the 100 simulations are 1.9$\times 10^{-2}$ 
for \psfent, 2.9$\times 10^{-2}$ for 5th-order polynomial models and 4.7$\times 10^{-2}$ 
for 5$\times$5 boxcar smoothing. 
That is, \psfent provides an improvement of $\sim 35\%$ over polynomials and $\sim 60\%$ over 
boxcar smoothing in interpolating the PSF size. Note that the $\Delta$ value for polynomial fitting appears 
much larger than that derived in H12. This is because the atmospheric variations in our 100 images is much 
larger than the data used in H12, where they have used 33 images, all taken within 2 nights. We conclude 
here that \psfent is capable of interpolating other general shape parameters, once we properly informed the 
priors from simulations. }

\section{Correlation between ellipticity components}
\label{sec:e1e2_corr}

\chihway{We have argued in \Sref{sec:likelihood} that the two ellipticity components 
should be independent from each other in each exposure, while both varying with similar 
amplitudes and spatial structures between different exposures. In this appendix we 
demonstrate this by calculating the following three spatial correlation functions for the 
100 ``true'' PSF images:}
\begin{equation}
 \xi_{11}(\theta)=\langle \varepsilon_{1}(\theta_{0}) \varepsilon_{1} (\theta_{0}+\theta) \rangle;
 \label{eq:cf_diagnostic1}
\end{equation}
\begin{equation}
 \xi_{22}(\theta)=\langle \varepsilon_{2}(\theta_{0}) \varepsilon_{2} (\theta_{0}+\theta) \rangle;
 \label{eq:cf_diagnostic2}
\end{equation}
\begin{equation}
 \xi_{12}(\theta)=\langle \varepsilon_{1}(\theta_{0}) \varepsilon_{2} (\theta_{0}+\theta) \rangle.
 \label{eq:cf_diagnostic3}
\end{equation}

\begin{figure}
   \subfigure{\includegraphics[scale=0.35]{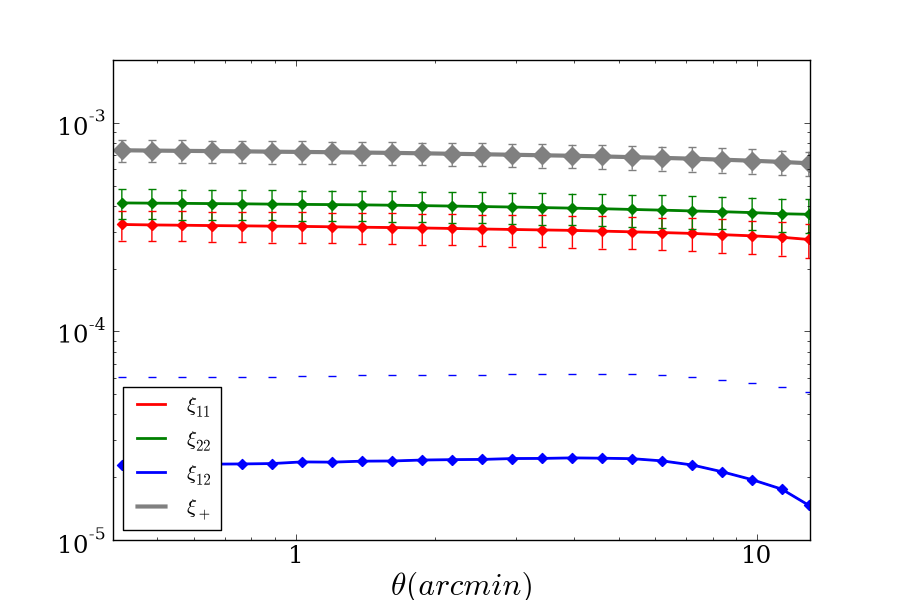} }
 \caption{\chihway{Three diagnostic correlation functions $\xi_{11}$ (red), $\xi_{22}$ (green) and 
 $\xi_{12}$ (blue), plotted with the full ellipticity correlation function $\xi_{+}$ (grey). Each 
 curve represents the median of the 100 different atmosphere realisations. The error bars 
 indicate the rms spread in the 100 exposures divided by $\sqrt{100}$.}}
 \label{fig:diagnostic}
\end{figure}

\chihway{In \Fref{fig:diagnostic}, the three correlation functions are shown, each representing 
the median of the 100 realisations in the sample, with the error bars showing the standard 
deviation of the 100 realisations divided by $\sqrt{100}$. The $\xi_{11}$ and $\xi_{22}$ 
curves are clearly positive and show similar structures. The difference in the two curves may 
be due to sample variance, and the fact that there are no rotation/dithering between these 
exposures. On the other hand, $\xi_{12}$ is consistent with zero. This supporting our 
argument that the ``phase'' of the two ellipticity components are independent of each other, 
while the spatial power spectrum is similar.}

\end{document}